\documentclass[11pt,letterpaper]{article}
\usepackage[margin=1in]{geometry}
\usepackage{rpmacros}
\usepackage[T1]{fontenc}
\usepackage{gitinfo}
\usepackage[usenames,dvipsnames]{xcolor}
\usepackage{stmaryrd}
\usepackage{xfrac}
\usepackage{mathpazo}
\usepackage{bm}
\usepackage{todonotes}
\usepackage{lipsum}
\usepackage{setspace}
\usepackage{scrextend}
\usepackage[shortlabels]{enumitem}
\usetikzlibrary{matrix}

\usepackage{calrsfs}
\DeclareMathAlphabet{\pazocal}{OMS}{zplm}{m}{n}
\RequirePackage[colorlinks=true]{hyperref}
\hypersetup{
	linkcolor=[rgb]{0.3,0.3,0.6},
	citecolor=[rgb]{0.2, 0.6, 0.2},
	urlcolor=[rgb]{0.6, 0.2, 0.2}
}
\usepackage{amsthm}
\usepackage{thmtools}
\usepackage{thm-restate}

\numberwithin{equation}{section}
\declaretheoremstyle[bodyfont=\it,qed=\qedsymbol]{noproofstyle}

\declaretheorem[numberlike=equation]{observation}

\declaretheorem[name=Observation,numbered=no]{observation*}

\declaretheorem[numberlike=equation]{theorem}
\declaretheorem[numberlike=equation,style=noproofstyle,name=Theorem]{theoremwp}
\declaretheorem[name=Theorem,numbered=no]{theorem*}

\declaretheorem[numberlike=equation]{lemma}
\declaretheorem[name=Lemma,numbered=no]{lemma*}
\declaretheorem[numberlike=equation,style=noproofstyle,name=Lemma]{lemmawp}

\declaretheorem[numberlike=equation]{corollary}
\declaretheorem[name=Corollary,numbered=no]{corollary*}

\declaretheorem[name=Proposition,numbered=no]{proposition*}
\declaretheorem[numberlike=equation,style=noproofstyle,name=Proposition]{propositionwp}

\declaretheorem[numberlike=equation]{claim}
\declaretheorem[name=Claim,numbered=no]{claim*}
\declaretheorem[numberlike=equation,style=noproofstyle,name=Claim]{claimwp}

\declaretheorem[name=Conjecture,numbered=no]{conjecture*}

\declaretheorem[numberlike=equation]{question}
\declaretheorem[name=Question,numbered=no]{question*}

\declaretheoremstyle[bodyfont=\it,qed=$\lozenge$]{defstyle} 

\declaretheorem[numberlike=equation,style=defstyle]{definition}
\declaretheorem[unnumbered,name=Definition,style=defstyle]{definition*}

\declaretheorem[unnumbered,name=Example,style=defstyle]{example*}

\declaretheorem[unnumbered,name=Notation=defstyle]{notation*}

\declaretheorem[unnumbered,name=Construction,style=defstyle]{construction*}

\declaretheorem[numberlike=equation,style=defstyle]{remark}
\declaretheorem[unnumbered,name=Remark,style=defstyle]{remark*}

\renewcommand{\phi}{\varphi}
\renewcommand{\epsilon}{\varepsilon}
\newcommand{\K}{\mathbb{K}}

\newcommand{\size}{\operatorname{size}}
\newcommand{\coeff}{\operatorname{coeff}}
\newcommand{\eval}{\operatorname{eval}}

\newcommand{\vecZ}{\mathbf{Z}}

\newcommand{\OR}{\operatorname{OR}}

\newcommand{\Perm}{\operatorname{Perm}}
\newcommand{\cvector}{\overline{\operatorname{coeff}}}

\newcommand{\C}{\mathbb{C}}

\newcommand{\RSDesign}{\operatorname{RS-Design}}
\newcommand{\g}{\mathsf{Gen}}

\newcommand{\inbracket}[1]{\llbracket #1 \rrbracket}

\newcommand{\Mon}{\operatorname{Mon}}
\newcommand{\Bin}{\operatorname{Bin}}
\newcommand{\Mod}{\operatorname{Mod}}

\newcommand{\calp}{\pazocal{P}}
\newcommand{\cktSize}{\operatorname{CircuitSize}}
\newcommand{\expSumSize}{\operatorname{ExpSumSize}}
\newcommand{\ckt}{\operatorname{Circuit}}
\newcommand{\expSum}{\operatorname{ExpSum}}

\newcommand{\class}[1]{\mathcal{#1}}

\newcommand{\Burgisser}{B\"{u}rgisser}

\usepackage{nth}
\usepackage{intcalc}
\usepackage{etoolbox}
\usepackage{xstring}

\usepackage{ifpdf}
\ifpdf
\else
\usepackage[quadpoints=false]{hypdvips}
\fi

\newcommand{\ECCCurl}[2]{http://eccc.hpi-web.de/report/\ifnumcomp{#1}{>}{93}{19}{20}#1/#2/}

\newcommand{\shortECCC}[2]{\texttt{\href{http://eccc.hpi-web.de/report/\ifnumcomp{#1}{>}{93}{19}{20}#1/#2/}{eccc:TR#1-#2}}}

\newcommand{\parseECCC}[1]{% Takes a string of the form TRxx/xxx or
\StrSubstitute{#1}{TR}{}[\tmpstring]%
\IfSubStr{\tmpstring}{/}{ %assuming string is of the form TRxx/xxx
\StrBefore{\tmpstring}{/}[\ecccyear]%
\StrBehind{\tmpstring}{/}[\ecccreport]%
}{% assuming string is of the form TRxx-xxx
\StrBefore{\tmpstring}{-}[\ecccyear]%
\StrBehind{\tmpstring}{-}[\ecccreport]%
}%
\shortECCC{\ecccyear}{\ecccreport}}

\newcommand{\ignore}[1]{}

\newif\ifnote
\notetrue
\ifnote
\newcommand{\RPnote}[1]{\textcolor{BrickRed}{\guillemotleft RP: #1 \guillemotright}}
\newcommand{\ATnote}[1]{\textcolor{OliveGreen}{\guillemotleft AT: #1 \guillemotright}}
\newcommand{\PCnote}[1]{\textcolor{NavyBlue}{\guillemotleft PC: #1 \guillemotright}}
\newcommand{\CRnote}[1]{\textcolor{Purple}{\guillemotleft CR: #1 \guillemotright}}
\newcommand{\MKnote}[1]{\textcolor{Orange}{\guillemotleft MK: #1 \guillemotright}}
\else
\newcommand{\RPnote}[1]{}
\newcommand{\ATnote}[1]{}
\newcommand{\PCnote}[1]{}
\newcommand{\CRnote}[1]{}
\newcommand{\MKnote}[1]{}
\fi

\newcommand{\ehref}[1]{\href{mailto:#1}{#1}}

\allowdisplaybreaks
\onehalfspace
 
\title{On the Existence of Algebraic Natural Proofs\footnote{This paper builds on a combination of two preliminary works, titled ``On the Existence of Algebraically Natural Proofs'' (FOCS 2020) and ``If VNP is hard, then so are equations for it'' (STACS 2022).}}
\author{
	Prerona Chatterjee\thanks{\ehref{prerona.ch@gmail.com}. Parts of this work was done as a postdoc at Tel Aviv University, Israel (supported by the Azrieli International Postdoctoral Fellowship and ISF grant number 514/20) and as a PhD student at STCS, Tata Institute of Fundamental Research, Mumbai, India.}
  \and
  Mrinal Kumar\thanks{\ehref{mrinal.kumar@tifr.res.in}. School of Technology and Computer Science, TIFR, Mumbai, India. Most of this work was done while working at the Department of Computer Science \& Engineering, IIT Bombay. }
  \and
	C. Ramya\thanks{\ehref{ramyac@imsc.res.in}. The Institute of Mathematical Sciences (a CI of Homi Bhabha National Institute), Chennai, India. A part of this work was done while the author was a post-doctoral fellow at TIFR, Mumbai, India.}
  \and
	Ramprasad Saptharishi\thanks{\ehref{ramprasad@tifr.res.in}. School of Technology and Computer Science, Tata Institute of Fundamental Research, Mumbai, India. Research supported by Ramanujan Fellowship of DST, and by DAE, Government of India.}
  \and
	Anamay Tengse\thanks{\ehref{anamay.tengse@gmail.com}. School of Computer Sciences, NISER. Part of this work was done as a postdoc at Reichman University, Herzliya, supported by ISF grant 843/23, and as a PhD student at STCS, TIFR, Mumbai, India.}%
}

\date{}

\begin{document}
	
\maketitle

% {\let\thefootnote\relax
% \footnotetext{\textcolor{white}{(France's longest land border is with Brazil.) Base version:~(\gitAuthorIsoDate)\;,\;\gitAbbrevHash\;\; \gitVtag}}
% }

\begin{abstract}

  The framework of \emph{algebraically natural proofs} was independently introduced in the works of Forbes, Shpilka and Volk (2018), and Grochow, Kumar, Saks and Saraf (2017), to study the efficacy of commonly used techniques for proving lower bounds in algebraic complexity.
  We use the known connections between algebraic hardness and pseudorandomness to shed some more light on the question relating to this framework, as follows.
  \begin{itemize}
    
    \item The subclass of $\VP$ that contains polynomial families with bounded coefficients, \emph{has} efficient equations.
    Over finite fields, this result holds without any restriction on coefficients.
    Further, both these results extend to \emph{any} class that admits a low-variate, low-degree universal map: a generator for all polynomials in the class.
    Most well-studied classes have this property, e.g. $\VNP$, $\mathsf{VBP}$, $\mathsf{VF}$.

    \item Over fields of characteristic zero, $\VNP$ \emph{does not have} any efficient equations, if the Permanent is exponentially hard for algebraic circuits. Moreover, exponential hardness of the Permanent in the approximative sense, even rules out efficient equations of large degree.\\
    This gives the only known barrier to ``natural'' lower bound techniques (that follows from believable hardness assumptions), and also shows that the restriction on coefficients in the first category of results about $\VNP$ is necessary.
  
  \end{itemize}

The first set of results is obtained by algebraizing the well-known method of generating hardness from non-trivial hitting sets, and by generalizing the result of Heintz and Schnorr (1980) that proves the existence of hitting sets for $\VP$.
The conditional hardness of equations for $\VNP$ uses the fact that pseudorandomness against a class can be extracted from a polynomial that is (sufficiently) hard for that class (Kabanets and Impagliazzo, 2004).

% We strongly believe that despite the simplicity of their proofs, our results underscore the peculiar nature of the questions surrounding algebraic natural proofs, making them interesting even outside the context of proving algebraic circuit lower bounds. 

\end{abstract}

\newpage

\section{Introduction}\label{sec:introduction}

The quest for proving strong lower bounds for algebraic circuits is one of the fundamental challenges in algebraic complexity, and maybe the most well-studied one.
Yet, progress on this problem has been painfully slow and sporadic.
Perhaps the only thing more frustrating than the inability to prove such lower bounds is the difficulty in coming up with plausible approaches towards them.
This lack of progress has spurred an interest towards understanding the viability of some commonly used lower bound approaches; the idea being that a good sense of what approaches will \emph{not} work would aid in the search of those that \emph{might} work.
Moreover, such meta-studies could help identify the strengths of the current and future approaches that show promise.

In the broader context of lower bounds in computational complexity, there are various results of this flavor which establish that various families of techniques cannot be used for proving very strong lower bounds. 
For instance, the barrier of \emph{Relativization} due to Baker, Gill and Solovay \cite{BGS75}, that of \emph{Algebraization} due to Aaronson and Wigderson \cite{AW09} and that of \emph{Natural Proofs} due to Razborov and Rudich \cite{RR97}.\footnote{Sometimes, these results are conditional, as in \cite{RR97}.}
While none of these barrier results are directly applicable to the setting of algebraic computation, there have been recent attempts towards generalizing these ideas to the algebraic set up. A key notion in this line of work is the notion of \emph{algebraically natural proofs} alluded to and defined  in the works of Aaronson and Drucker \cite{AD08}, Forbes, Shpilka and Volk \cite{FSV18}, and Grochow, Kumar, Saks and Saraf  \cite{GKSS17}.

We now discuss this notion, starting with a discussion on \emph{Natural Proofs} which motivated the definition.

\subsection{The Natural Proofs framework of Razborov and Rudich} 
Razborov and Rudich \cite{RR97} noticed that underlying many of the  lower bound proofs known in Boolean circuit complexity,  there was some common structure. They formalized this common structure via the notion of a \emph{Natural Property}, which we now define. 
\begin{definition}\label{def:natural prop}
A subset  ${\pazocal{P}} \subseteq \{f: \{0,1\}^n \to \{0,1\}\}$ of Boolean functions is said to be a natural property useful against a class $\class{C}$ of Boolean circuits if the following are true. 
\begin{itemize}
\item \textbf{Usefulness.} Any Boolean function $f:\{0,1\}^n \to \{0,1\}$ that can be computed by a Boolean circuit in $\class{C}$ does not have the property $\pazocal{P}$.  
\item \textbf{Constructibility.} Given the truth table of a Boolean function $f: \{0,1\}^n \to \{0,1\}$, whether it has the property $\pazocal{P}$ can be decided in time polynomial in the length of the input, i.e. in time $2^{O(n)}$. 
\item \textbf{Largeness.} For all large enough $n$, at least a $2^{-O(n)}$ fraction of all $n$ variate Boolean functions have the property ${\pazocal{P}}$. \qedhere
\end{itemize}
\end{definition}

A proof that a certain family of Boolean functions cannot be computed by circuits in $\class{C}$ is said to be a \emph{natural lower bound proof} if it (perhaps implicitly) proceeds via establishing a natural property useful against $\class{C}$, and showing that the candidate hard function has this property.
Razborov and Rudich then showed that most of the Boolean circuit lower bound proofs that we know (for example, lower bounds for $\mathsf{AC}^0$ circuits \cite{FSS84, H86} or lower bounds for $\mathsf{AC}^0[\oplus]$ circuits \cite{R87, S87}) fit into this framework, maybe with some work, and hence are \emph{natural} in this sense.
Further they argue that, under standard cryptographic assumptions, the proof of a lower bound against any sufficiently rich circuit class (such as the class $\P/\poly$) cannot be natural!
Thus, under standard cryptographic assumptions, most of the current lower bound techniques are not strong enough to show super-polynomial lower bounds for general Boolean circuits.

We now move on to discuss a relatively recent analogue of the notion of Natural Proofs, formalized in the context of algebraic computation.

\subsection{Algebraically Natural Proofs}

Algebraic complexity is the study of computational questions about polynomials as formal objects.
The basic model of computation here, an algebraic circuit, is an  algebraic analogue of a boolean circuit with the gates of the circuit being labeled by $+$ (sum) and $\times$ (product) gates as opposed to Boolean functions; the size of a circuit is the number of wires (edges) in it.\footnote{See \autoref{defn:alg-circuits} for a formal definition. }
The algebraic analogue of $\P/\poly$ is the class $\VP$ of polynomial families $\{f_n\}$, where $f_n$ is an $n$ variate polynomial of degree and algebraic circuit size $\poly(n)$.
A fundamental  question in this setting is to come up with explicit families of polynomials, those in the class $\VNP$ (the algebraic analog of $\NP/\poly$), which are not in $\VP$.
While the state of the art of size lower bounds for algebraic circuits is a bit better than that for Boolean circuits, with slightly super linear lower bounds having been shown by Strassen~\cite{S73} and Baur \& Strassen~\cite{BS83}, this lower bound has seen no improvements for nearly four decades.
The recent breakthrough by Limaye, Srinivasan and Tavenas~\cite{LST24} does indeed suggest that stronger lower bounds might be within reach in the near future, but it is far from clear how that could be done.
This absence of progress has led to some research towards understanding the limitations of the current proof techniques in proving strong lower bounds for algebraic circuits. 

Considering that algebraic circuits seem like a fairly general and powerful model of computation, it is tempting to think that the \emph{natural proofs barrier} of Razborov and Rudich \cite{RR97} also extends to this setting.
However, this problem turns out to be a non-trivial one, and indeed, it is not known whether their results extend to algebraic circuits.
This question is closely related to the existence of cryptographically secure, algebraic pseudorandom functions that can be computed by small and low degree\footnote{Throughout this paper, by a \emph{low degree} polynomial family, we mean a polynomial family whose degree is polynomially bounded in its number of variables.} algebraic circuits, and there does not seem to be substantial evidence one way or the other on this.
We refer the reader to \cite{AD08} and  \cite{FSV18} for a more detailed discussion on this issue.

In the last few years, this question of trying to find an algebraic analogue of the barrier results in \cite{RR97} has received substantial attention.
It was observed by various authors \cite{AD08, G15, FSV18, GKSS17} that most of the currently known  proofs of algebraic circuit lower bounds fit into a  common unifying framework, not unlike that in \cite{RR97}, although of a more algebraic nature.
Indeed,  these proofs also implicitly go via defining a property for the set of all polynomials and using this property to separate the hard polynomial from the easy ones.
Moreover, the notions of \emph{largeness}  and \emph{constructibility} in \autoref{def:natural prop} also seem to extend to these proofs. 

We now discuss this framework in a bit more detail. The key notion here is that of an equation for a set of polynomials. 

\begin{restatable}[Equations for a set of polynomials]{definition}{EquationsForSet}\label{def:proofs-set-of-polynomials}
	For some $n, d \in \N$, let $\pazocal{C}_{n,d}$ be a set of $n$-variate polynomials of \emph{total} degree at most $d$; i.e. $\pazocal{C}_{n,d} \subseteq \F[\vecx]^{\leq d}$.
	
	Then, for $N = \binom{n + d}{n}$, a nonzero polynomial $P_N(\vecZ)$ is said to be an \emph{equation} for {$\pazocal{C}_{n,d}$} if for all $f(\vecx) \in \pazocal{C}_{n,d}$, we have that $P_{N}(\cvector(f)) = 0$, where $\cvector(f)$ is the coefficient vector of $f$.
\end{restatable}

The definition naturally extends to a class of polynomial families, as opposed to just a set of polynomials as defined above. In particular, suppose that $\class{C}$ is a class of polynomial families $\inbrace{\{f_{n}\} : f_{n} \in \pazocal{ C}_{n,d_n}}$, and  $\{P_N\}$ is a polynomial family.
Then, the family $\{P_N\}$ is said to be a family of equations for {$\class{C}$} if $P_{N(n)}$ is an equation for  $\pazocal{C}_{n,d_n}$ for all large enough $n$, for $N(n) := \binom{n+d_n}{n}$.
That is, there is some $n_0$ such that for all $n \geq n_0$ the polynomial $P_{N(n)}$ is an equation for  $\pazocal{C}_{n,d_n}$. 

Intuitively, non-vanishing of an equation (for a set $\pazocal{C}$) on the coefficient vector of a given polynomial $f$ is a proof that $f$ is not in $\pazocal{C}$.
We note that the equations for a set $\pazocal{C}$ evaluate to zero not just on the coefficient vectors of polynomials in $\pazocal{C}$ but also on the coefficient vectors of polynomials in the Zariski closure of $\pazocal{C}$.
This framework comes up very naturally in the context of algebraic geometry (and geometric complexity theory), where it is often geometrically nicer to work with the variety obtained by taking the Zariski closure of a complexity class.

Getting our hands on an equation of a variety gives us a plausible way to test and certify non-membership in the variety, in other words, to prove a lower bound for the corresponding complexity class.
Thus, families of equations for a class gives an algebraic analogue of the notion of \emph{natural properties useful against a class} in \cite{RR97}.
Moreover, since a nonzero polynomial does not vanish very often on a random input from a large enough grid, it follows that a nonzero equation for a set $\pazocal{C}$ will be nonzero on the coefficient vector of a ``random polynomial''.
Here by a random polynomial we mean a polynomial whose coefficients are independent and uniformly random elements from some large enough set in the underlying field.
With appropriate quantitative bounds, this observation can be formalized to give an appropriate algebraic analogue of the notion of \emph{largeness}.
Lastly, the algebraic circuit complexity of the equation gives a natural algebraic analog of the notion of \emph{constructibility}.
Intuitively, any algebraic circuit lower bound which goes via defining a nonzero proof polynomial of polynomially bounded degree that can be efficiently computed by an algebraic circuit is an \emph{Algebraically Natural Proof} of a lower bound. 

We now formally define an algebraically natural proof. 
\begin{definition}[Algebraically natural proofs \cite{FSV18, GKSS17}]
	Let $\class{C}$ be a class of polynomial families $\inbrace{\{f_{n,d}\} : f_{n,d} \in \pazocal{C}_{n,d}}$. 
  
	Then, for a class $\class{D}$ of polynomial families,  we say that \emph{$\class{C}$ has $\class{D}$-natural proofs}  if there is a family $\{P_N\} \in \class{D}$ which is a non-trivial family of equations for  $\class{C}$. 
\end{definition}
In the rest of this paper, whenever we say a natural proof, without specifying the class $\class{D}$, we mean a $\VP$-natural proof. 

Analogous to the abstraction of \emph{natural proofs} for Boolean circuit lower bounds, this framework of \emph{algebraically natural proofs} turns out to be rich and general enough that  almost all of our current proofs of algebraic circuit lower bounds are in fact algebraically natural, or can be viewed in this framework with a little work \cite{G15}.
Thus, this definition seems like an important first step towards understanding the strengths and limitations of many of our current lower bound techniques in algebraic complexity. 

The immediate next question to ask is whether  algebraically natural proofs are rich enough to give strong algebraic circuit lower bounds.
This can naturally be worded in terms of the complexity of equations for the class $\VP$ as follows.   
\begin{question}\label{q:main}
	For every constant $c > 0$, does there exist a nonzero polynomial family $\{P_{N, c}\}$ in $\VP$ such that for all large enough $n$, the following is true?  
	\begin{quote}
		For every family of polynomials $\{ f_n \}$ in $\VP$, such that $f_n$ is an $n$ variate polynomial of degree $n^c$, $P^{(c)}_{N}$ vanishes on the coefficient vector of $f_n$ for $N = \binom{n + n^c}{n}$. 
	\end{quote}
\end{question}

The works \cite{FSV18} and \cite{GKSS17} argue that under an appropriate (but non-standard) pseudorandomness assumption, the answer to the question above is negative, i.e., algebraically natural proof techniques cannot be used to show strong lower bounds for algebraic circuits.
To discuss this pseudorandomness assumption formally, we need to define \emph{succinct hitting sets}.

\begin{definition}[Succinct hitting sets for a set of polynomials]
  	For some $n, d \in \N$, let $\pazocal{C}_{n,d}$ be a set of $n$-variate polynomials of \emph{total} degree at most $d$; that is, $\pazocal{C}_{n,d} \subseteq \F[\vecx]^{\leq d}$.
  
  	Then for $N = \binom{n + d}{n}$, we say that a set of $N$ variate polynomials \emph{$\pazocal{D}_N$ has $\pazocal{C}_{n,d}$-succinct hitting sets} if for all nonzero $P(\vecZ) \in \pazocal{D}_{N}$, there exists some $f \in \pazocal{C}_{n,d}$ such that $P_{N}(\cvector(f)) \neq 0$.
\end{definition}

As with \autoref{def:proofs-set-of-polynomials}, this definition naturally extends to polynomial families (see \autoref{defn:succinct-hitting-sets}). 

It immediately follows from these definitions that non-existence of $\class{D}$-natural proofs against a class $\class{C}$ is equivalent to the existence of $\class{C}$-succinct hitting sets for the class $\class{D}$.
Forbes, Shpilka and Volk \cite{FSV18} showed that for various restricted circuit classes $\class{C}$ and $\class{D}$, the class $\class{D}$ has $\class{C}$ succinct hitting sets.
Or equivalently, lower bounds for $\class{C}$ cannot be proved via proof polynomial families in $\class{D}$.
However, we already have super-polynomial lower bounds against these classes $\class{C}$, making the evidence weak.
Further, this question has remained unanswered for more general circuit classes $\class{C}$ and $\class{D}$.
In particular, if we take both $\class{C}$ and $\class{D}$ to be $\VP$, we do not seem to have significant evidence on the existence of $\VP$-succinct hitting sets\footnote{The definition of $\VP$-succinct hitting sets (\autoref{defn:vp-succinct-hitting-sets-for-vp}) is perhaps slightly non-intuitive.} for $\VP$.

In \cite{FSV18}, the authors observed that showing $\VP$ succinct hitting sets for $\VP$ would immediately imply non-trivial deterministic algorithms for polynomial identity testing which, via well known connections between algebraic hardness and derandomization, will in turn imply new lower bounds \cite{HS80a, KI04}.
Thus, the problem of proving an unconditional barrier result for algebraically natural proof techniques via this route seems as hard as proving new circuit lower bounds! 
It is, however, conceivable that one can show such a barrier conditionally.
And in some more structured settings, such as for the case of matrix completion, such results are indeed known \cite{BIJL18}.
However, \autoref{q:main} remains open.  In particular,  even though many of the structured subclasses of $\VP$ have low degree equations which are very efficiently computable, perhaps hoping that this extends to richer and more general  circuit classes is too much to ask for? 

\subsection{Our results}

We are now ready to state our results.
Our first set of results can be viewed as evidence \emph{towards} the efficacy of natural techniques for proving lower bounds against $\VP$, and possibly even $\VNP$.

\subsubsection*{Equations for polynomials in $\VP$ with coefficients of small complexity}

We first show that over the field of complex numbers, there are efficiently computable equations for the set of polynomials in $\VP$ that have small coefficients.
Here for a field $\F$, $\VP^{\F}$ denotes the class $\VP$ where the coefficients are from the field $\F$.

\begin{restatable}{theorem}{MainThmComplex}\label{thm:complexes}
	Let $c > 0$ be any constant. 
  There is a polynomial family $\{P^{(c)}_{N}\} \in \VP^{\Q}$ such that for $N(n) = \binom{n+n^c}{n}$, the following are true. 
  \begin{itemize}\itemsep0pt
  	\item  For every $t(n) = \poly(n)$, for all large enough $n$, and every family $\{f_n\} \in \VP^{\C}$, where $f_n$ is an $n$-variate, degree-$n^c$, size-$t(n)$ polynomial with coefficients in $\{-1, 0, 1\}$, we have that
  	\[
   		P^{(c)}_{N}(\cvector(f_n)) = 0 \, ,  
  	\] 
  	where $\cvector(f_n)$ is the coefficient vector of $f_n$. 
  	\item There exists a family $\set{h_n}$ of $n$ variate polynomials and degree $\leq n^c$ with coefficients in $\{-1, 0, 1\}$ such that for all large enough $n$,
  	\[
  		P^{(c)}_{N}(\cvector(h_n)) \neq 0 \, .
   	\]
  \end{itemize}
\end{restatable}
We note that even though \autoref{thm:complexes} is stated for polynomials with $\{-1, 0, 1\}$ coefficients, the theorem holds for polynomials with coefficients as large as $N$.

However, for brevity, we will confine the discussion in this paper to polynomials with coefficients in $\{-1, 0, 1\}$.
We also note that the same statement holds over other fields of characteristic zero as well.
That is, the $\VP^{\C}$ in the above statement can be replaced with $\VP^{\R}$ or $\VP^{\Q}$.

We also prove an analogous theorem for finite fields. 

\begin{restatable}{theorem}{MainThmFiniteFields}\label{thm:finite field}
 	Let $\F$ be any finite field, and let $c > 0$ be any constant.
  There is a polynomial family $\set{P^{(c)}_{N}} \in \VP^{\F}$ such that for $N(n) = \binom{n+n^c}{n}$, the following are true. 
	\begin{itemize}\itemsep0pt
	\item  For every $t(n) = \poly(n)$, for all large enough $n$, and every family $\{f_n\} \in \VP^{\F}$, where $f_n$ is an $n$-variate, degree-$n^c$, size-$t(n)$ polynomial, we have that
  	\[
    	P^{(c)}_{N}(\overline{\coeff}(f_n)) = 0 \, .
  	\]
	\item There exists a family $\set{h_n}$ of $n$ variate polynomials and degree $\leq n^c$ with coefficients in $\F$ such that for all large enough $n$,
	\[
		P^{(c)}_{N}(\overline{\coeff}(h_n)) \neq 0 \, .
	\]
	\end{itemize}  
\end{restatable}

\subsubsection*{Equations for polynomials in $\VNP$ with coefficients of small complexity}

Furthermore, we also prove analogous statements for the seemingly larger class $\VNP$, as follows.

\begin{restatable}{theorem}{VNPboundedCoeff}\label{thm:vnp-complexes}
	Let $c > 0$ be any constant. 
	There is a polynomial family $\{Q^{(c)}_{N}\} \in \VP^{\Q}$ such that for $N(n) = \binom{n+n^c}{n}$, the following are true. 
	\begin{itemize}\itemsep0pt
		\item  For every $t(n) = \poly(n)$, for all large enough $n$, and every family $\{f_n\} \in \VNP^{\C}$, where $f_n$ is an $n$-variate, degree-$n^c$, size-$t(n)$ polynomial with coefficients in $\{-1, 0, 1\}$, we have that
	  \[
	   	Q^{(c)}_{N}(\cvector(f_n)) = 0 \, .  
	  \] 
  	\item There exists a family $\set{h_n}$ of $n$ variate polynomials and $\leq n^c$ with coefficients in $\{-1, 0, 1\}$ such that for all large $n$,
  	\[
  		Q^{(c)}_{N}(\cvector(h_n)) \neq 0 \, .
    \]
  \end{itemize}
\end{restatable}

As before, we note that the above theorem holds for polynomials with coefficients as large as $N$ and also holds over other fields of characteristic zero, like $\R$ or $\Q$.

\begin{restatable}{theorem}{VNPfiniteField}\label{thm:vnp-ff}
	Let $\F$ be any finite field and $c > 0$ be any constant.
  There is a polynomial family $\set{Q^{(c)}_{N}} \in \VP^{\F}$ such that for $N(n) = \binom{n+n^c}{n}$, the following are true. 
	\begin{itemize}
		\item  For every $t(n) = \poly(n)$, for all large enough $n$, and every family $\{f_n\} \in \VNP^{\F}$, where $f_n$ is an $n$-variate, degree-$n^c$, size-$t(n)$ polynomial, we have that
		\[
			Q^{(c)}_{N}(\overline{\coeff}(f_n)) = 0 \, .
		\]
		\item There exists a family $\set{h_n}$ of $n$ variate  polynomials and degree $\leq n^c$ with coefficients in $\F$ such that for all large $n$,
		\[
			Q^{(c)}_{N}(\overline{\coeff}(h_n)) \neq 0 \, .
		\]
	\end{itemize}  
\end{restatable}

In fact, we show that the existence of efficient hitting sets for any class is sufficient to give efficient equations for its subclass that contains polynomials with bounded coefficients.
This is formalized in \autoref{thm:equations-from-hitting-sets-char0} for fields of characteristic zero, and in \autoref{thm:equations-from-hitting-sets-finite-fields} for finite fields.

\subsubsection*{Conditional hardness of equations for \textsf{VNP}}

Over fields of characteristic zero, we also show that assuming the Permanent is hard enough, the constraint of bounded coefficients in \autoref{thm:vnp-complexes} is necessary for efficient equations for $\VNP$.
More formally, we show the following.

\begin{restatable}[Conditional Hardness of Equations for {\sf VNP}]{theorem}{NoNaturalProofsVNP}\label{thm:vnp-equations-hard}
  Let $\epsilon > 0$ be a constant.
  Suppose that the permanent family $\Perm_m$ requires circuits of size $2^{m^{\epsilon}}$. 
  
  Then, $\VP$ has $\VNP$-succinct hitting sets.
  Therefore, there are no $\VP$-natural proofs for $\VNP$. 
\end{restatable}

We remark that the above theorem holds over any field of characteristic zero.
For our proofs, we will work with complexes for better readability.

\begin{remark}
  Extending the result in \autoref{thm:vnp-equations-hard} to hardness of equations for $\VP$, even under the assumption that Permanent is sufficiently hard, is a fascinating open question.
  Such an extension would answer the main question investigated in \cite{FSV18, GKSS17} and show a natural-proofs-like barrier for a fairly general family of lower bound proof techniques in algebraic complexity.
  Our proof of \autoref{thm:vnp-equations-hard} however, crucially relies on some key properties of $\VNP$, and does not appear to extend to $\VP$.
\end{remark}

\subsubsection*{Conditional approximative hardness of equations for \textsf{VNP}}

Finally, we show that the statement can be strengthened further if we assume that ${\Perm_n}$ is hard even in the approximative sense.

\begin{definition}[Approximative circuits and complexity]\label{defn:approximative-circuits-complexity}
  A circuit $C(\vecx,\epsilon)$ is said to \emph{approximately compute} an $n$-variate polynomial $f(\vecx)$ if it satisfies $C(\vecx,\epsilon) = \epsilon^{M} \cdot f(\vecx) + \epsilon^{M+1} \cdot g(\vecx,\epsilon) \in \F(\epsilon)[\vecx]$ for some $(n+1)$-variate polynomial $g$, and $M \in \N$.

  For a polynomial $f(\vecx)$, its \emph{approximative complexity} is defined to be the size of the smallest circuit that approximately computes it.
\end{definition}

It is clear that the approximative complexity of $f$ is at most $\cktSize(f)$, just because one can take $M$ to be $0$, and $g(\vecx,\epsilon)$ to be the identically zero polynomial in \autoref{defn:approximative-circuits-complexity}.

We show that if we assume that ${\Perm_n}$ is exponentially hard even in the approximative sense, then we can rule out all efficient equations for $\VNP$, irrespective of their degree.
More formally, we show the following.

\begin{restatable}[Conditional Approximative Hardness of Equations for {\sf VNP}]{theorem}{NoEfficientEquationsForVNP}\label{thm:vnp-equations-hard-border}
  Let $\epsilon > 0$ be a constant.
  Suppose that the permanent family $\Perm_m$ requires approximative circuits of size $2^{m^{\epsilon}}$.
  
  Then $\VP_{\mathrm{nb}}$ has $\VNP$-succinct hitting sets, and so there are no efficiently computable equations for $\VNP$. 
\end{restatable}

This is an important statement, especially when we think of degree as a measure of the ``largeness'' of the purported equation.

\subsection{Discussion and relations to prior work}

As is evident from our results, the main message (in our opinion) is that we do not have compelling evidence to rule out, or accept, the efficacy of algebraically natural proofs towards proving strong lower bounds for rich classes of algebraic circuits.
In fact, our results seem to provide \emph{some} evidence for both sides.

We first discuss the results that suggest an affirmative answer to \autoref{q:main}.
Many of the families of polynomials commonly studied in algebraic complexity have integer coefficients with absolute values bounded by $1$, and fall in the setting of \autoref{thm:complexes}.
Moreover, the condition of computing polynomials with bounded coefficients is a semantic condition on a model, in the sense that even though the final output of the circuit is required to have bounded coefficients, the circuit is free to use arbitrary constants from $\C$ in the intermediate computation.
Thus, it is conceivable that we might be able to prove a super-polynomial lower bound on the algebraic circuit size for the permanent polynomial via an algebraically natural proof constructible in $\VP$, thereby separating $\VP$ and $\VNP$.
However, since analogues of \autoref{thm:complexes} and \autoref{thm:finite field} are also true for $\VNP$, any such separation will have to rely on more fine-grained information on the equations, and not just their degree and algebraic circuit size. 
Unfortunately, our proofs are all existential and do not give a sense of what the polynomial families $\{P^{(c)}_{N}\}$ (or $\{Q^{(c)}_{N}\}$) might look like.

We also note that in the light of some prior works, these results are perhaps a bit surprising.
The classes of polynomials in $\VP$ and $\VNP$ with small coefficients (or over finite fields), are seemingly rich and complex, and the theorems here show --- unconditionally --- that they have equations which are also efficiently computable.
Furthermore, these equations are a rather straightforward consequence of the existence of efficient hitting sets, as shown in Theorems \ref{thm:equations-from-hitting-sets-finite-fields} and \ref{thm:equations-from-hitting-sets-char0}.
As discussed earlier, the existence of efficient equations is known to be true for many structured subclasses of algebraic circuits (for example, homogeneous  circuits of depth $3$ and $4$, multilinear formulas, polynomials of small Waring rank).
However, it is unclear if this property extends to more general circuit classes, like $\VP$ or $\VNP$.

\paragraph*{} % this is to add a space between paragraphs

On the other hand, \autoref{thm:vnp-equations-hard} shows that under the widely believed assumption that $\Perm$ is exponentially hard, the bound on coefficients is crucial, at least for the class $\VNP$.
It is unclear if a similar situation is also true for $\VP$.
However, even if we were to believe that there are efficiently computable equations for $\VP$, it is unclear if the existence of such equations implies $\VP \neq \VNP$.

In the other direction, suppose  we were to assume that $\VP$ and $\VNP$ are indeed different, is it then reasonable to expect an efficiently computable equation exhibiting such a separation?
Using \autoref{thm:vnp-equations-hard}, we now have that if $\Perm$ is exponentially hard, then  \emph{any} efficiently computable equation for $\VP$ will necessarily \emph{not} be an equation for $\VNP$, thus yielding an ``algebraically natural proof''  that separates  $\VP$ and $\VNP$.

\paragraph*{} % this is to add a space between paragraphs

We will now briefly go over some works related to \autoref{q:main}.
As mentioned earlier, the focus of most of the prior works has been to look for evidence that the answer to \autoref{q:main} is negative, i.e. $\VP$ does not have efficiently computable and low degree equations.
We hope that the results in this paper highlight that the answer is not so clear.

\paragraph*{Relations to prior work.}
Following the work of Forbes, Shpilka, Volk \cite{FSV18} and Grochow, Kumar, Saks and Saraf \cite{GKSS17}, much of the research on the problem, of whether algebraically natural proofs exist, has focused on proving the \emph{non-existence} of efficiently computable equations for $\VP$, and this line of work has made interesting progress in this direction for many structured and special instances of problems of this nature.
Forbes, Shpilka and Volk \cite{FSV18} unconditionally ruled out equations for depth-three multilinear formulas computable by certain structured classes of algebraic circuits using this connection. However, this does not imply anything about complexity of equations for general classes of algebraic circuits such as $\VP$ and $\VNP$.

In the context of proving lower bounds against algebraic circuits, Efremenko, Garg, Oliveira and Wigderson \cite{EGOW18} and Garg, Makam, Oliveira and Wigderson \cite{GMOW19} explore limitations of proving algebraic circuit lower bounds via rank based methods.
In particular, Efremenko \etal~\cite{EGOW18} show that some of these rank based methods cannot prove lower bounds better than $\Omega_d(n^{\lfloor d/2 \rfloor})$ on tensor rank (respectively, Waring rank) for a $d$-dimensional tensor of side $n$.
Building on \cite{EGOW18}, in \cite{GMOW19}, the authors demonstrate that one \emph{cannot} hope to significantly improve the known lower bounds for tensor rank for $d$ dimensional tensors  by lifting lower bounds on tensors in fewer dimensions.
However,  we note that a general algebraically natural proof of a lower bound does  not necessarily fit into the framework of \cite{EGOW18, GMOW19}, and so these limitations for the so-called \emph{rank methods} do not seem to immediately extend to algebraically natural proofs in general.
As discussed earlier, in the light of the results here, it is conceivable that we might be able to improve the state of the art for general algebraic circuit lower bounds, using techniques that are algebraically natural. 

Bl\"{a}ser, Ikenmeyer, Jindal and Lysikov~\cite{BIJL18} studied the complexity of equations in a slightly different context.
They draw connections between the existence of efficiently constructible equations of a variety and the problem of testing (non)membership in it and use the conditional hardness of the (non)membership testing problem for certain varieties to rule out the existence of efficiently computable equations for them.
More precisely, they show that if \emph{all} the equations for the variety of matrices with zero permanent are constructible by small constant-free algebraic circuits, then  the  non-membership problem for this variety can be decided in the class $\exists \BPP$.
Thus, unless $\P^{\#\P} \subseteq \exists \BPP$, the equations of this variety do not have small, low degree constant free algebraic circuits.
In a subsequent work (\cite{BILPS19}), the results of \cite{BIJL18} are generalized  to  \emph{min-rank} or \emph{slice-rank} varieties.
However, in the bounded coefficient setting (and over finite fields), our results show that the contrary is true, and $\VP$ does have efficiently computable low degree equations.
We also note that the set-up in these papers differ from that in our paper, and that of \cite{GKSS17,FSV18}.
One way to interpret this difference is that \cite{BIJL18} shows that ``variety of small completion rank tensors'' cannot be ``cut out'' by efficient equations, whereas \cite{GKSS17, FSV18} and our paper asks if \emph{every} equation for this variety requires large complexity.  

A positive result on the complexity of equations of naturally-occurring varieties in algebraic complexity appears in a recent work of Kumar and Volk \cite{KV20} where they show polynomial bounds on the degree of the equations of the Zariski closure of the set of non-rigid matrices and small linear circuits over all large enough fields.
However, we do not know if any of these low-degree equations can be \emph{efficiently} computed by an algebraic circuit. 

For Boolean circuits, Chow \cite{C11} shows a way of circumventing the natural proofs' barrier in  \cite{RR97} by providing (under standard cryptographic assumptions) an explicit \emph{almost natural proof} that is useful against $P/\poly$ as well as constructive in nearly linear time, but compromises on the largeness condition.
Furthermore, Chow \cite{C11} shows the unconditional existence of a natural property useful against $P/\poly$ (infinitely often) constructive in linear size that has a weakened largeness condition.
In some sense, \autoref{thm:finite field} and \autoref{thm:complexes} are analogous to the work of Chow \cite{C11}, albeit in the algebraic world.

\paragraph{On the largeness criterion.} In the definitions of algebraically natural proofs \cite{GKSS17, FSV18}, the authors observe that in the algebraic setting, an analogue of the \emph{largeness} criterion in \autoref{def:natural prop} is often available for free; the reason being that a nonzero equation for any class of polynomials  vanishes on a very small fraction of all polynomials over any sufficiently large field.
However, this tradeoff becomes a bit subtle when considering polynomials over finite fields, or polynomials with bounded integer coefficients.
In particular, as we observe in the course of the proofs of our results, we still have a sizeable chunk of polynomials whose coefficients will keep $\{P^{(c)}_{N}\}$ (and $\{Q^{(c)}_{N}\}$) nonzero, although this set is no longer a significant fraction of the set of all polynomials.

\subsection{Proof ideas}

\subsubsection*{Constructible equations for polynomials with coefficients of small complexity}

At a high level, the idea behind our results about constructible equations is to try and come up with a \emph{non-trivial} property of polynomials which every polynomial with a small circuit satisfies. 
By a non-trivial property, we mean that there should exist (nonzero) polynomials which do not have this property.
The hope is that once we have such a property (that is nice enough), one can try to transform this into an equation via an appropriate \emph{algebraization}.
The property that we finally end up using is the existence of \emph{hitting sets} for polynomials with small circuits.

A hitting set for a class $\class{C}$ of polynomials over a field $\F$ is a set of points $\pazocal{H}$, such that every nonzero polynomial in $\pazocal{C}$ evaluates to a nonzero value on at least one point in $\pazocal{H}$. We then turn this property of \emph{not-vanishing-everywhere on $\pazocal{H}$} into an equation in some settings. In order to formalize this, let us consider the map $\Phi_{\pazocal{H}}$ defined on the set of all polynomials, using the hitting set $\pazocal{H}$ of a class $\class{C}$, that maps any given polynomial $f$ to its evaluations over the points in $\pazocal{H}$.
It is clear from the above observation that any nonzero polynomial in the kernel of $\Phi_{\pazocal{H}}$ is guaranteed to be outside $\class{C}$.
Thus, if there were a nonzero polynomial that vanishes on all polynomials $f \notin \operatorname{ker}(\Phi_{\pazocal{H}})$, then we would have an equation for $\class{C}$.

Moreover, if such a polynomial happened to have its degree and circuit complexity polynomially bounded in its number of variables, we would have our required upper bounds.
However, note that \emph{not} being in the kernel of a linear map seems to be a tricky condition to check via a polynomial (as opposed to the complementary property of \emph{being} in the kernel, which can be easily checked via a polynomial).
To prove our theorems, we get past this issue in the setting of finite fields, and for polynomials over $\C$ with bounded integer coefficients.

Over a finite field $\F$, a univariate polynomial that maps every nonzero $x \in \F$ to zero and vice versa, already exists in $q(x) = 1 - x^{\abs{\F} - 1}$.
Therefore, for a given polynomial $f$, the equation essentially outputs $ \prod_{\vech \in \pazocal{H}} q(f(\vech))$.
Clearly, for a polynomial $f$, $ \prod_{\vech \in \pazocal{H}} q(f(\vech))$ is zero if and only if $f$ evaluates to a nonzero value on at least one point in $\pazocal{H}$. 
To generalize this to other fields, we wish to find a ``low-degree'' univariate $q(x)$ that maps nonzero values to 0, and zero to a nonzero value.
We observe when the polynomials in $\class{C}$ have integer coefficients of bounded magnitude we can still obtain such a univariate polynomial, and in turn a non-trivial equation.
In particular, if $q(\vecx)$ were such a univariate, we essentially output $ \prod_{\vech \in \pazocal{H}} q(f(\vech))$, for a given polynomial $f$.
This step relies on a simple application of the Chinese Remainder Theorem. 

In order to show that the equations are non-trivial, in the sense that there exist polynomials with bounded integer coefficients which do not pass this test, we need to show that there are nonzero  polynomials with bounded integer coefficients which vanish everywhere on the hitting set $\pazocal{H}$.
We show this via a well-known lemma of Siegel\footnote{A statement of the lemma can be found \href{https://en.wikipedia.org/wiki/Siegel\%27s_lemma}{here}.
Refer to \cite{S14a} for details.}, which uses a simple pigeon-hole argument to show that an under-determined system of homogeneous linear equations where the constraint matrix has small integer entries has a nonzero solution with small integer entries. 

As it turns out,  our proofs do not use much about the class $\VP$ except for the existence of small hitting sets for polynomials in the class (\autoref{thm:equations-from-hitting-sets-char0}).
In fact, even the existence of these hitting sets essentially follows universal circuits or polynomials, for $\VP$ (\autoref{thm:equations-from-universal-maps-char0}). 
It is not hard to observe that these properties are also true for the seemingly larger class $\VNP$ and hence the results here also extend to $\VNP$.
We also note that, given the hitting set $\pazocal{H}$ explicitly, the construction of the equation is completely explicit.
In other words, the non-explicitness in our construction comes only from the fact that we do not have explicit constructions of hitting sets for algebraic circuits.

\subsubsection*{$\VNP$-succinct hitting sets for $\VP$}

As was observed in \cite{FSV18, GKSS17}, a lower bound for equations for a class of polynomials is equivalent to showing the existence of succinctly describable hitting sets for this class.
For our proof of \autoref{thm:vnp-equations-hard} we show that, assuming that the permanent is sufficiently hard, the coefficient vectors of polynomials in $\VNP$ form a \emph{hitting set} for the class $\VP$.
The connection between hardness and randomness in algebraic complexity is well known via a result of Kabanets and Impagliazzo~\cite{KI04}, and we use this connection for our proof, along with some additional ideas.
It is useful to note that the above-mentioned result of Kabanets and Impagliazzo~\cite{KI04} is essentially an algebraic analogue of the \emph{hardness vs randomness} paradigm introduced by Nisan and Wigderson~\cite{NW94} in the boolean world.
We briefly describe a high level sketch of our proof in a bit more detail now.

Kabanets and Impagliazzo~\cite{KI04} showed that using any explicit polynomial family $\{f_n\}$ that is sufficiently hard, one can construct (a family of) hitting set generators for $\VP$.
That is, we can construct a polynomial map $\g_f:\F^{k} \rightarrow \F^{t}$ that ``fools'' any small algebraic circuit $C$ on $t$ variables in the sense that $C(y_1, y_2, \ldots, y_t)$ is nonzero if and only if the $k$-variate polynomial $C\circ\g_f$ is nonzero.
In a typical invocation of this result, the parameter $k$ is much smaller than $t$ (typically $k = \poly\log t$).
Thus, this gives a reduction from the question of polynomial identity testing for $t$-variate polynomials to polynomial identity testing for $k$-variate polynomials.
Another related way of interpreting this connection is that if $\{f_n\}$ is sufficiently hard then $\g_f$ is a polynomial map whose image does not have an equation with small circuit size.
Thus, assuming the hardness of the Permanent, this immediately gives us a polynomial map (with appropriate parameters) such that its image does not have an efficiently constructible equation.

For the proof of \autoref{thm:vnp-equations-hard} (more formally, \autoref{thm:lb-on-equations-for-vnp-slice}), we show that the points in the image of the map $\g_{\Perm}$, can be viewed as the coefficient vectors of polynomials in $\VNP$, or, equivalently in the terminology in \cite{FSV18, GKSS17}, that the Kabanets-Impagliazzo hitting set generator is $\VNP$-succinct.
To this end, we work with a specific instantiation of the construction of the Kabanets-Impagliazzo generator where the underlying construction of combinatorial designs is based on Reed-Solomon codes.
Although this is perhaps the most well known construction of combinatorial designs, there are other (and in some parameters, better) constructions known.
However, our proof relies on the properties of this particular construction to obtain the succinct description.
Our final proof is fairly short and elementary, and is based on extremely simple algebraic ideas and making generous use of the fact that we are trying to prove a lower bound for equations for $\VNP$ and not $\VP$.

For \autoref{thm:vnp-equations-hard-border}, we prove an approximative version of \autoref{thm:lb-on-equations-for-vnp-slice} by using the result of {\Burgisser}~\cite{B18}\footnote{The \href{https://doi.org/10.1007/s10208-002-0059-5}{original work} is from 2004, this version corrects an error in one of the proofs.} which showed that approximative circuits are particularly useful in computing factors of polynomials that have small size, irrespective of their degree.

\paragraph*{On algebraic ``PRFs'' against $\VP$:} As said above, the main proof can be summarized as saying that the Kabanets-Impagliazzo generator~\cite{KI04} applied on the symbolic permanent is $\VNP$-succinct. Informally, this states that if the symbolic permanent is exponentially hard, then the coefficient vectors of polynomials in $\VNP$ ``look random'' to polynomials in $\VP$. If the succinctness of this (or any other) generator can be improved to $\VP$, then this would be a definitive step towards completely ruling out the existence of efficiently computable equations for $\VP$. 

\subsection{Organization of the paper. } 
We begin with some notations and preliminaries in \autoref{sec:prelims}.
In \autoref{sec:hitting-sets}, we outline the existence of efficient hitting sets, over characteristic-zero fields, for any class that has a low-degree, low-variate \emph{universal polynomials}; $\VP$ and $\VNP$ are examples of such classes.
Existence of efficient hitting sets for $\VP$ and $\VNP$ over finite fields is outlined in \autoref{subsec:hitting-sets-finite-fields}.
Note that all the arguments in \autoref{sec:hitting-sets} are present in previous works, but the results over characteristic zero have not been stated in this generality prior to this work, to the best of our knowledge.
In \autoref{sec:equations}, we use the above results to first prove Theorems \ref{thm:complexes} to \ref{thm:vnp-ff}.
We then show the conditional hardness of equations for $\VNP$ in \autoref{sec:vnp-succinct-hitting-sets-for-vp}.
Finally, we provide some open questions that arise from our work, and the prior literature around algebraic natural proofs in \autoref{sec:open-questions}.

\section{Notation and preliminaries}\label{sec:prelims}

\subsection{Notation and basics}

\begin{itemize}
	\item We use $[n]$ to denote the set $\set{1,\ldots, n}$ and $\inbracket{n}$ to denote the set $\set{0,1,\ldots, n}$. We also use $\N_{\geq 0}$ to denote the set of non-negative integers.

	\item As usual, we identify the elements of $\F_p$ with $\{0, 1, \ldots, p-1\}$ and think of $\inbracket{n}$ as a subset of $\F_p$ in the natural way for any $n < p$.

	\item We use boldface letters such as $\vecx, \vecy$ to denote tuples, typically of variables. When necessary, we adorn them with a subscript such as $\vecy_{[n]}$ to denote the length of the tuple. We also use $\vecx^{\vece}$ to denote the monomial $\prod x_i^{e_i}$.

	\item We use $\{f_n\}_{n \in \N}$ to denote families of polynomials. We drop the index set whenever it is clear from context. For a given polynomial $f$ we denote by $\deg(f)$ its degree. For a polynomial $f(\vecx,\vecy,\ldots)$ on multiple sets of variables, we use $\deg_{\vecx}(f) $, $\deg_{\vecy}(f) $, etc., to denote the degree in the variables from the respective sets.

	\item We use $\F[\vecx]^{\leq d} $ to denote polynomials over the field $\F$ in variables $\vecx$ of degree at most $d$, and use $\vecx^{\leq d} $ to denote the set of all monomials in variables $\vecx$ of degree at most $d$.

	\item For a given polynomial $f \in \F[\vecx]^{\leq d} $ and a monomial $m \in \vecx^{\leq d} $, we use $\coeff_{m}(f)$ to refer to the coefficient of $m$ in $f$. We further use $\cvector(f)$ to denote the vector\footnote{We do not explicitly mention the monomial ordering used for this vector representation, since all our statements work for any monomial ordering.} of coefficients of $f$.
	
  \item We denote sets of polynomials and classes of polynomial families by two sets of calligraphic letters. Sets are denoted by $\pazocal{C}$, $\pazocal{D}$, etc., and classes of families are denoted by $\class{C}$, $\class{D}$, etc.
\end{itemize}

We will require the notions of hitting sets and hitting set generators (HSGs) given below.

\begin{definition}[Hitting Set]\label{defn:hitting-set}
  A set of points $\pazocal{H}$ is said to be a hitting set for a set of polynomials $\pazocal{T}$, if for each $f \in T$ there exists an $h \in \pazocal{H}$ for which $f(h) \neq 0$. 
\end{definition}

\begin{definition}[Hitting Set Generator (HSG)]\label{defn:hitting-set-generator}
  A vector of polynomials $(g_1(\vecy),\ldots,g_n(\vecy))$ is said to be a \emph{hitting set generator} for a set of $n$-variate polynomials $\pazocal{T}$, if for each $f \in \pazocal{T}$ we have that the composed polynomial $f(g_1(\vecy),\ldots,g_n(\vecy))$ is nonzero.
\end{definition}

\noindent We will also be using the well-known Polynomial Identity Lemma.

\begin{lemma}[Polynomial Identity Lemma~\cite{O22,DL78,S80,Z79}]\label{lem:polynomial-identity}
Let $f\in \F[x_1, x_2, \ldots, x_n]$ be a nonzero polynomial of degree at most $d$ and let $S$ be a subset of $\F$ (or an extension of $\F$). Then, the number of zeroes of $f$ on the grid $S^n$ is at most $d\abs{S}^{n-1}$.  
\end{lemma}

\begin{corollary}\label{cor:trivial-hitting-set-PIL}
For $S = \set{0,1,\ldots,d}$, the grid $S^n$ is a hitting set for the set of all $n$-variate polynomials of degree at most $d$.
\end{corollary}

\subsection{Algebraic circuits and complexity classes}

Let us first formally define algebraic circuits.

\begin{definition}[Algebraic circuits]\label{defn:alg-circuits} 
	An \emph{algebraic circuit} is specified by a directed acyclic graph, with leaves (in-degree zero; also called \emph{inputs}) labelled by field constants or variables, and internal nodes labelled by $+$ or $\times$. The  nodes with out-degree zero are called the \emph{outputs} of the circuit. Computation proceeds in the natural way, where inductively each $+$ gate computes the sum of its children and each $\times$ gate computes the product of its children.

  The \emph{size} of the circuit is defined as the number of nodes in the underlying graph. 
\end{definition}

\begin{definition}[Exponential sums of circuits]\label{defn:exponential-sums}
  For an algebraic circuit $C(\vecx,\vecy)$, the \emph{exponential sum of $C$ over $\vecy$} is defined as follows.
  \[
    \widetilde{C}(\vecx) := \sum_{\alpha \in \set{0,1}^{\abs{\vecy}}} C(\vecx,\vecy = \alpha)
  \]
  The \emph{size} of such an exponential sum is said to be the sum of the size of $C$, and the number of ``auxiliary'' variables $\vecy$.
  Similarly, the \emph{degree} of the exponential sum is simply the total degree (in $\vecx$ and $\vecy$) of the polynomial computed by $C$.
  For instance, the exponential sum $\widetilde{C}$ has size equal to $\size(C) + \abs{\vecy}$, and its degree is the total degree of $C(\vecx,\vecy)$.
\end{definition}

\begin{definition}[Size of a polynomial]\label{defn:size-of-polynomial}
  For a polynomial $f(\vecx)$, the \emph{circuit-size of $f$} is defined as the size of the smallest circuit that computes it.
  It is denoted as $\cktSize(f)$.

  Similarly, the \emph{exponential-sum-size of $f$} is defined as the size of the smallest exponential sum that computes it.
  This is denoted by $\expSumSize(f)$.
\end{definition}

\subsubsection{Polynomial families and their complexity}

In order to study the asymptotic cost of computing polynomials, we work with the families of polynomials that they naturally define (e.g. $\set{\det_n}$ is the family of $n \times n$ determinants for all natural $n$).
We formally define polynomial families and their complexity for clarity. 

\begin{definition}[Polynomial families]\label{defn:polynomial-families}
  For a set of variables $\vecx$ and a field $\F$, a family of polynomials within $\F[\vecx]$, denoted by $\set{f_n}_{\N} $ (or simply $\set{f_n}$), is a set of polynomials indexed by $n \in \N$ such that the $n$-th polynomial $f_n$ depends on at most $n$ variables for all $n \in \N$.

  We shall denote the collection of all such polynomial families using $\calp$. 
\end{definition}

\begin{remark}\label{rmk:polynomial-families}
It is more common to work with what are called `$p$-bounded families', where the number of variables that the polynomials in the family $\set{f_n}$ depend on, grows polynomially with the index $n$.
We choose the stricter definition above for clarity, and brevity of our statements.
More importantly, all the contents of this paper can be translated to the language of $p$-bounded families.
\end{remark}

The main focus of this work is going to be families of polynomials in which the degree grows polynomially with the number of variables, called \emph{low-degree polynomial families}.

\begin{definition}[Low-degree polynomial families]\label{defn:low-degree-families}
  For a function $d : \N \rightarrow \N$, we define the class of all degree-$d$ polynomial families as follows.
  \[
        \class{P}_d := \set{ \set{f_n} \in \class{P} : \forall n, \deg(f_n) \leq d(n) }
  \]

  The collection of \emph{low-degree polynomial families} is then naturally defined as the union of $\class{P}_d $'s over all polynomial functions $d$.
  \[
        \class{P}_{\mathrm{low-deg}} := \bigcup_{c \in \N} \class{P}_{n^c}. \qedhere
  \]
\end{definition}

Note that every family in $\class{P}_{\mathrm{low-deg}}$ implicitly fixes a constant $c > 0$ so that the family belongs to the class $\class{P}_d$ where $d:n \mapsto n^c$.
A helpful feature of $\class{P}_d$ is that for each $\set{f_n}$ in $\class{P}_d$, the length of the \emph{coefficient vector} of the $n^{th}$ polynomial $f_n$ is exactly $\binom{n+d(n)}{n}$ and therefore is \emph{purely a function of $n$}.
This makes it easier to handle equations for sets of polynomials, which is the central subject of our work.
Therefore, we will always work with specific subclasses of $\class{P}_d$ for a fixed $d$ while dealing with equations, and this choice of $d$ will be mentioned explicitly whenever it is not completely clear from the context.

\subsubsection*{Complexity classes of polynomial families}

We can now define classes of low-degree families that are efficiently expressible using circuits and exponential sums.

\begin{definition}[Computable families]\label{defn:computable-families}
  For functions $s : \N \rightarrow \N$ and $d : \N \rightarrow \N$, we define the class of degree-$d$ families that are computable by size $s$ circuits as follows.
  \[
        \ckt_{d,s} := \set{ \set{f_n} \in \class{P}_d : \exists n_0 \in \N, \forall n > n_0, \cktSize(f_n) \leq s(n) }
  \]
  When $d$ and $s$ are polynomials, we shall alternatively refer to this class as $\VP_{d,s}$.
\end{definition}

\begin{definition}[Definable families]\label{defn:definable-families}
  For functions $s : \N \rightarrow \N$ and $d : \N \rightarrow \N$, we define the class of degree-$d$ families that are expressible by exponential sums of size and degree $s$, as follows.
  \[
        \expSum_{d,s} := \set{ \set{f_n} \in \class{P}_d : \exists n_0 \in \N, \forall n > n_0, \expSumSize(f_n) \leq s(n) } \qedhere
  \]
  For polynomials $d,s$, we shall sometimes refer to this class as $\VNP_{d,s}$.
\end{definition}

We now define the familiar classes $\VP$ and $\VNP$ more formally, specifically to help the forthcoming discussion about families of equations for these classes.

\begin{definition}[\textsf{VP} from first principles]\label{defn:VP-first-principles}
  For a polynomially bounded $d : \N \rightarrow \N$, we denote by $\VP_d$, the class of all degree-$d$ polynomial families that are efficiently computable using algebraic circuits, as follows.
  \[
     \VP_d := \union_{e \in \N} \VP_{d,n^{e}} = \union_{e \in \N} \ckt_{d,n^{e}} 
  \]

  The collection of all efficiently computable low-degree polynomial families denoted by $\VP$, is then naturally defined as follows.
  \[  \VP :=  \union_{c \in \N} \VP_{n^{c}} \qedhere \] 
\end{definition}

\begin{definition}[\textsf{VNP} from first principles]\label{defn:VNP}
  For a polynomially bounded $d : \N \rightarrow \N$, we denote by $\VNP_d$, the class of all degree-$d$ polynomial families that are efficiently definable using exponential sums, as follows.
  \[
     \VNP_d := \union_{e \in \N} \VNP_{d,n^{e}} = \union_{e \in \N} \expSum_{d,n^{e}}
  \]

  The collection of all efficiently definable low-degree polynomial families denoted by $\VNP$, is then naturally defined as follows.
  \[  \VNP :=  \union_{c \in \N} \VNP_{n^{c}} \qedhere \] 
\end{definition}

\paragraph*{Dependence on the field.} All the definitions discussed so far can be naturally instantiated for any field $\F$.
We give a short summary for clarity.
\begin{itemize}
  \item A polynomial family $\set{f_n}$ over a field $\F$ is a family in which the coefficients of $f_n$ are elements in $\F$ for all $n$.\\
  Similarly, we can then define $\class{P}_d^{\F}$ to be the class of degree-$d$ families over $\F$.
  \item A circuit is said to be over a field $\F$, if all the constants appearing in the circuit are from $\F$.
  Exponential sums over $\F$ are defined similarly.
  \item We then define $\ckt_{d,s}^{\F}$ and $\expSum_{d,s}^{\F}$ to be the analogous subclasses of families over the field $\F$, according to circuits and exponential sums over $\F$.
  \item Finally, $\VP^{\F}$ is defined as the union over all $c,e \in \N$, of the classes $\ckt_{d,s}^{\F}$, with $d = n^c$ and $s = n^e$.\\
  Similarly, $\VNP^{\F}$ is the union of all $\expSum_{d,s}^{\F}$.
\end{itemize}

An important point to note here is that $\VP^{\F}$ and $\VNP^{\F}$ are defined for a \emph{fixed} field $\F$.
Particularly, in the case of finite fields, the size of the field is a constant with respect to $n$.
The underlying field will usually be clear from the context, and will therefore not be mentioned unless required.

We will now formally define algebraic natural proofs and succinct hitting sets over characteristic zero.
Defining these for finite fields adds one more subtlety of largeness, which we address after that.

\subsubsection{Equations, natural proofs and succinct hitting sets over characteristic zero}

To start with, we recall the definition of equations for a set of polynomials.

\EquationsForSet*

Observe that any class $\class{C} \subseteq \class{P}_d$ naturally defines a set of $n$-variate, degree-$d(n)$ polynomials: $\class{C}(n) = \set{ f_n : \set{f_n} \in \class{C} }$.
We shall use this piece of notation in order to define the concept of natural proofs for a class of polynomial families.

\begin{definition}[Natural proofs for a class of families]\label{defn:natural-proofs}
  For some functions $d,D : \N \rightarrow \N$, and classes $\class{C} \subseteq \class{P}_d$, $\class{D} \subseteq \class{P}_D $, we say that a family $\set{P_N}$ is a \emph{$\class{D}$-natural proof for $\class{C}$}, if:
  \begin{itemize}\itemsep0pt
    \item $\set{P_N} \in \class{D}$, and
    \item for all large enough $n$, $P_N$ is an equation for $\class{C}(n)$ for $N = \binom{n+d(n)}{n}$. \qedhere
  \end{itemize}
\end{definition}

As first defined in the works of Forbes, Shpilka and Volk \cite{FSV18}, and Grochow, Kumar, Saks and Saraf \cite{GKSS17}, we then get the following definition for `succinct hitting sets' by essentially negating the definition of natural proofs.

\begin{definition}[Succinct hitting sets for a class of families]\label{defn:succinct-hitting-sets}
  For functions $d,D : \N \rightarrow \N$, and classes $\class{C} \subseteq \class{P}_d$, $\class{D} \subseteq \class{P}_D $, we say that \emph{$\class{C}$-succinct hitting sets exist for $\class{D}$}, if the following is true.
  \begin{quote}
    \emph{For infinitely many} $n \in \N$, the set of coefficient vectors of $\class{C}(n)$ is a hitting set for the set of polynomials $\class{D}(N)$ where $N = \binom{n+d(n)}{n}$. \qedhere
  \end{quote}
\end{definition}

The following statement is therefore an immediate consequence of the definitions above.
\begin{propositionwp}\label{prop:natural-proofs-opp-succinct-hs}
  Over any field of characteristic zero, for classes $\class{C} \in \class{P}_d$ and $\class{D} \in \class{P}_D$, $\class{C}$ has $\class{D}$-natural proofs if and only if $\class{D}$ does not have $\class{C}$-succinct hitting sets.
\end{propositionwp}

\paragraph*{Handling $\VP$, $\VNP$.}
We now instantiate the definitions of natural proofs and succinct hitting sets for the specific cases of $\VP$ and $\VNP$.
This needs a bit of care because both $\VP$ and $\VNP$ are collections of countably many classes of polynomial families, each containing polynomials families of a specific degree.
Thus, a formal definition of say, ``$\VP$-natural proofs for $\VP$'', does not directly follow from \autoref{defn:natural-proofs}, and one has to rely on the purpose of defining the concept, which is that of analyzing the power of known techniques in the context of proving non-membership in $\VP$.
This leads us to the following definitions.

\begin{definition}[$\VP$-natural proofs for $\VP$]\label{defn:vp-natural-proofs-for-vp}
  
  For polynomially bounded functions $d,s,D,S : \N \rightarrow \N$, we say that a family $\set{P_N}$ is a \emph{$\VP_{D,S}$-natural proof for $\VP_{d,s}$}, if:
  \begin{itemize}\itemsep0pt
    \item $\set{P_N} \in \VP_{D,S}$, and
    \item for all large enough $n$, $P_N$ is an equation for $\VP_{d,s}(n)$ for $N = \binom{n+d(n)}{n}$.
  \end{itemize}

  We say that \emph{$\VP$-natural proofs exist for $\VP$} if the following is true.
  \begin{quote}
    For any $d(n) \in \poly(n)$, there exist $D(N)$, $S(N) \in \poly(N)$, and a family $\set{P_N}$, such that for every $s(n) \in \poly(n)$, $\set{P_N}$ is a $\VP_{D,S}$-natural proof for $\VP_{d,s}$.
  \end{quote}

  In other words, for every degree-function $d(n)$, there is a single family $\set{P^{(d)}_N} \in \VP$, that is a family of equations for all of $\VP_{d} = \cup_s \VP_{d,s}$.
\end{definition}

Two important aspects of this definition must be noted.
\begin{itemize}
  \item The proof family $\set{P_N}$ \emph{may depend} on the degree-function $d(n)$.
  
  If a polynomial $P_N$ vanishes on the coefficient vector of an $n$-variate degree-$d$ polynomial $f$, then the length of the coefficient vector should match the arity of the equation, and therefore $N = \binom{n+d}{d}$. Therefore, it is inevitable to see a dependence between $N$ and the degree-function $d(n)$. 
  % As mentioned earlier, this helps in establishing a clear relationship between $n$ and $N$.
  % But even beyond that, all the known instances of ``natural proofs for $\class{C}$'' have this property.
  % For instance, the partial derivative method applied to depth-3-powering circuits is captured by determinants of different sizes, depending on the degree of the underlying polynomial family.

  \item The proof family $\set{P_N}$ \emph{must not depend} on the size-function $s(n)$.
  
  This is crucial for a proof that yields a super-polynomial lower bound.
  That is, if the proof family were to depend on the size-function $s(n)$, it will only yield size-$s$ lower bounds: arbitrarily large polynomial lower bounds when $s(n)$ is a polynomial.
  
  That said, the ``prover'' is allowed to select a suitable ``largeness threshold $n_0$'' depending on the size-function $s(n)$.
  This is also easily justified.
  For instance, if $P_N$ vanishes on families that have circuits of size $\leq n^{\log n}$, then this $n_0$ would depend on when $n^{\log n}$ overshoots $s(n)$. 
\end{itemize}

We now define $\VP$-succinct hitting sets for $\VP$.
Similar to the general definition, this has been done so that ``existence of $\VP$-natural proofs for $\VP$'' and ``existence of $\VP$-succinct hitting sets for $\VP$'' are direct logical negations of each other.

\begin{definition}[$\VP$-Succinct hitting sets for $\VP$]\label{defn:vp-succinct-hitting-sets-for-vp}
  
  For polynomially bounded functions $d,s,D,S : \N \rightarrow \N$, we say that \emph{$\VP_{d,s}$-succinct hitting sets exist for $\VP_{D,S}$}, if the following is true.
  \begin{quote}
    For \emph{infinitely many} $n \in \N$, if $N = \binom{n+d(n)}{n}$ then the coefficient vectors of $\VP_{d,s}(n)$ form a hitting set for $\VP_{D,S}(N)$.
  \end{quote} 

  We say that \emph{$\VP$ has $\VP$-succinct hitting sets} if there exists a degree-function $d(n) \in \poly(n)$, such that for all choices of $D(N), S(N) \poly(N)$, there is a size-function $s(n) \in \poly(n)$, so that $\VP_{d,s}$-succinct hitting sets exist for $\VP_{D,S}$.
\end{definition}

\begin{propositionwp}
	$\VP$ has $\VP$-natural proofs $\Longleftrightarrow$ $\VP$ does not have $\VP$-succinct hitting sets.
\end{propositionwp}

\begin{remark}
  It can be argued that the term `$\VP$-succinct hitting sets' intuitively means that the hitting set can be described efficiently using algebraic circuits.
  This hints towards a fixed pair of degree and size functions working for all of $\VP$\footnote{``$\exists d(n), s(n) \forall D(N),S(N)$ \ldots'' as opposed to ``$\exists d(n) \forall D(N),S(N) \exists s(n) \ldots$''.}.
  Clearly, such a definition would be stronger than \autoref{defn:vp-succinct-hitting-sets-for-vp}.
  So a `succinct-hitting-sets-based-barrier' towards `lower bounds via natural proofs' in this intuitive sense is also a barrier as per \autoref{defn:vp-succinct-hitting-sets-for-vp}.
\end{remark}
  
Moreover, all the succinct hitting sets described in the work of Forbes, Shpilka and Volk \cite{FSV18} are consistent with the stronger, ``intuitive'' definition, and therefore also imply succinct hitting sets as per \autoref{defn:vp-succinct-hitting-sets-for-vp}.
More generally, as we shall see in \autoref{subsec:universal-circuits}, the notion of universal circuits lets us work with \autoref{defn:vp-succinct-hitting-sets-for-vp} with a suitable change in parameters.

\subsubsection{Algebraic natural proofs over finite fields}

As pointed out before, the classes $\VP^{\F}$ and $\VNP^{\F}$ are defined only for a fixed field $\F$, and hence all of $\VP_{d}(n)$ is a subset of $\F^N$.
We then have to rule out ``nonzero'' equations that vanish on the entire universe $\F^N$.

Taking a queue from the definition of (boolean) natural proofs (\autoref{def:natural prop}), we include the largeness criterion.
We first define the notion of a \emph{large polynomial family}.

\begin{definition}[Large polynomial families]\label{defn:large-polynomial-families-finite-fields}
  Let $\F = \F_q$ be the finite field of size $q$.
  For a constant $a$, a polynomial family $\set{P_N}$ over $\F$ is said to be \emph{$a$-large}, if for all large enough $N$, there is a set $W \subseteq \F^N$ of size at least $N^{-a} q^N$ such that $P_N(\vecw) \neq 0$ for all $\vecw \in W$.

  A family $\set{P_N}$ is said to be large, if it is $a$-large for some $a$.
\end{definition}

\begin{definition}[$\VP$-natural proofs for $\VP$ (Finite fields)]\label{defn:vp-natural-proofs-for-vp-finite-fields}
  
  Let $\F$ be a finite field.
  For polynomially bounded $d,s,D,S : \N \rightarrow \N$, we say that a family $\set{P_N}$ is a \emph{$\VP^{\F}_{D,S}$-natural proof for $\VP^{\F}_{d,s}$}, if:
  \begin{itemize}\itemsep0pt
    \item $\set{P_N} \in \VP^{\F}_{D,S}$,
    \item for all large enough $n$, $P_N$ is an equation for $\VP^{\F}_{d,s}(n)$ for $N = \binom{n+d(n)}{n}$, and
    \item $\set{P_N}$ is large as per \autoref{defn:large-polynomial-families-finite-fields}.
  \end{itemize}

  We say that \emph{$\VP^{\F}$-natural proofs exist for $\VP^{\F}$} if the following is true.
  \begin{quote}
    For any $d(n) \in \poly(n)$, there exists $D(N), S(N) \in \poly(N)$ and a family $\set{P_N}$, such that for every $s(n) \in \poly(n)$, $\set{P_N}$ is a $\VP^{\F}_{D,S}$-natural proof for $\VP^{\F}_{d,s}$.
  \end{quote}

  In other words, for each degree-function $d(n)$, there is a single family $\set{P^{(d)}_N} \in \VP^{\F}$, that is a family of equations for all of $\VP^{\F}_{d} = \cup_s \VP^{\F}_{d,s}$.
\end{definition}

By negating this definition, we get a seemingly weaker version of succinct hitting sets; in particular, the coefficient vectors are now only expected to hit the large families within $\VP^{\F}$.
Note that hitting all families in ``$\VP^{\F}(N)$'' is \emph{impossible} for ``$\VP^{\F}(n)$'', simply because the constant-degree, univariate polynomial $(Z_1^{q} - Z_1)$ vanishes over all of $\F_q^N$.

\begin{definition}[$\VP$-Succinct hitting sets for $\VP$ (Finite fields)]\label{defn:vp-succinct-hitting-sets-for-vp-finite-fields}
  
  Let $\F$ be a finite field.
  For polynomially bounded $d,s,D,S : \N \rightarrow \N$, we say that \emph{$\VP^{\F}_{d,s}$-succinct hitting sets exist for $\VP^{\F}_{D,S}$}, if the following is true.
  \begin{quote}
    For any large family $\set{P_N} \in \VP^{\F}_{D,S}$, for \emph{infinitely many} $n \in \N$, if $N = \binom{n+d(n)}{n}$ then the polynomial $P_N$ does not vanish on the coefficient vectors of $\VP_{d,s}(n)$.
  \end{quote} 

  We say that \emph{$\VP^{\F}$ has $\VP^{\F}$-succinct hitting sets} if there exists a degree-function $d(n) \in \poly(n)$ such that, for all choices of $D(N), S(N) \in \poly(N)$, there is a size-function $s(n) \in \poly(n)$ so that $\VP^{\F}_{d,s}$-succinct hitting sets exist for $\VP^{\F}_{D,S}$.
\end{definition}

\subsection{Universal Circuits}\label{subsec:universal-circuits}

A universal circuit is an algebraic circuit with the property that any polynomial which is efficiently computable is a simple projection of it.
The following lemma due to Raz shows the existence of such circuits; for the sake of completeness, we also include a proof sketch.

\begin{lemma}[Universal circuit~\cite{Raz10}]\label{lem:universal-circuit}
	Let $\mathbb{F}$ be any field and $n,s\geq 1$ and $d \geq 0$.
  Then there exists an algebraic circuit $U$ of size $\poly(n,d,s)$ computing a polynomial in $\mathbb{F}[x_1,\ldots,x_n,y_1,\ldots,y_r]$ with $r\leq \poly(n,d,s)$ such that:
	\begin{itemize}
		\item $\deg_{\vecx}(U(\vecx,\vecy)), \deg_{\vecy}(U(\vecx,\vecy)) \leq \poly(d)$;
		\item for any $f(\vecx) \in \ckt_{d,s}(n)$,  there exists an $\veca\in \mathbb{F}^r$ such that $f(\vecx)= U(\vecx,\veca)$.
	\end{itemize}
\end{lemma}

\begin{proof}
	Let $f$ be an $n$-variate degree $d$ polynomial computable by a circuit $C$ of size $s$. Using the classical depth reduction result due to Valiant \etal~\cite{VSBR83}, $f$ has a circuit $C'$ of size $s' = \poly(n,d,s)$ and depth $\ell = O(\log d)$ with the following properties (see, e.g., \cite{S15} for a complete proof).
	\begin{itemize}
	  	\item All the product gates have fan-in at most $5$.
	  	\item $C'$ is \emph{layered}, with alternating layers of sum and product gates.
	  	\item The layer above the leaves is of product gates, and the root is an addition gate.
	\end{itemize}
	
	We can therefore construct a \emph{layered} universal circuit $U$ for the given parameters $n,d,s$. The circuit will have $\ell$ layers, with $V_1, V_2, \ldots, V_{\ell} $ being the layers indexed from leaves to the root. So $V_{\ell}$ has a single gate, which is the output gate of the circuit, and $V_1$ has $n + 1$ gates, labeled with the variables $x_1,\ldots,x_n$ and with the constant $1$. All the gates in $U$ are then connected using auxiliary variables $\vecy$, as follows.
	\begin{itemize}
	  	\item $V_2$ has $\leq (n+1)^{5}$ product gates, with each gate computing a unique monomial of degree at most $5$ in the variables $\vecx$.
		\item For every odd $i$ with $2 < i < \ell$, the layer $V_i $ has $s'$ addition gates that are all connected to all the gates in the layer $V_{i-1}$, with each of the wires being labeled by a fresh $\vecy$-variable.
	  	\item For every even $i$ with $2 < i < \ell$, the layer $V_i$ has $\binom{s'}{5}$ product gates, each one multiplying a unique subset of $5$ gates from $V_{i-1}$.
	\end{itemize}
	
	It is now easy to see that $U$ has at most $\ell (n s')^5$ gates, which is $\poly(n,d,s)$. Also, $\deg(U) \leq 5^{\ell}$, which is $\poly(d)$; and $\abs{\vecy} = r \leq \ell \cdot (n s')^{6}$, which is $\poly(n,d,s)$.
  Further, by the depth reduction result~\cite{VSBR83}, the circuit $C'$ for $f$ can be obtained by setting the auxiliary variables $\vecy$ appropriately. Since the choice of $f$ was arbitrary, this finishes the proof.
\end{proof}

Universal circuits let us move naturally from succinct hitting sets to `succinct hitting set generators', which are defined as follows.
\begin{definition}[Succinct Hitting Set Generators]\label{defn:succinct-hitting-set-generators}
  Let $\F$ be any field and $n,d,N \in \N$ be such that $N = \binom{n+d}{n}$.
  For a set of $N$-variate polynomials $\pazocal{D}(N)$, a degree-$d$ polynomial $G_n \in \F[\vecy][x_1,\ldots,x_n]$ (seen as a polynomial only over the $\vecx$ variables) is said to be a \emph{succinct hitting set generator for $\pazocal{D}(N)$} if the coefficient vector of $G_n$ in ${(\F[\vecy])}^N$ is a hitting set generator for each polynomial $Q \in \pazocal{D}(N)$.

  Naturally, for a class of families $\class{D}$, a polynomial family $\set{G_n}$ is said to be a succinct hitting set generator (family) if, for infinitely many $n \in \N$, we have that $G_n$ is a succinct hitting set generator for $\pazocal{D}(N)$, where $n$ and $N$ are related according to the degree of $\set{G_n}$.

  When the family of generators $\set{G_n}$ belongs to a class $\class{C}$, we say that \emph{$\class{D}$ has a $\class{C}$-succinct hitting set generator}.
\end{definition}

\begin{lemma}[\cite{FSV18,GKSS17}]\label{lem:universal-succinct-hsg-for-classes}
For functions $d(n),s(n),D(N),S(N)$, if $\VP_{D,S}$ has $\VP_{d,s}$-succinct hitting sets then there is a family $\set{U_n}$, where $U_n$ is the universal circuit for parameters $n,d(n),s(n)$ guaranteed by \autoref{lem:universal-circuit}, that is a succinct hitting set generator for $\VP_{D,S}$.
\end{lemma}
\begin{proof}
  Let $\set{Q_N} \in \VP_{D,S}$ be a family.
  We will show that for infinitely many $N \in \N$, the composition $Q_N(\cvector(U_n))$ gives a nonzero polynomial in $\vecy$.
  
  Firstly, since $\VP_{D,S}$ has $\VP_{d,s}$-succinct hitting sets, there is some polynomial family ${f_n} \in \VP_{d,s}$ such that for infinitely many $N \in \N$, $Q_N(\cvector(f_n)) \neq 0$.
  From \autoref{lem:universal-circuit} we get that \emph{for all large enough} $n \in \N$, there is an assignment $\veca_n$ for which $U_n(\vecx,\vecy=\veca_n) = f_n(\vecx)$.
  
  Therefore, in particular, $Q_N(\cvector(U_n(\vecy=\veca_n))) = Q_N(\cvector(U_n))(\vecy=\veca_n) \neq 0$ and this implies that the composition $Q_N(\cvector(U_n))$ is a nonzero polynomial.
  This finishes the proof.
\end{proof}

\begin{lemma}\label{lem:succinct-hsg-for-vp}
Let $t(n) \in n^{\omega(1)}$ be any function.
If $\VP$ has $\VP$-succinct hitting sets, then there exists a polynomially bounded $d(n)$ such that the family of $n$-variate, degree-$d(n)$ universal circuits of size $t(n)$, forms a succinct hitting set generator for $\VP$.

As a result, $\VP$ has succinct hitting set generators of size $\poly(t(n))$.
\end{lemma}

\begin{proof}
Let us assume that $\VP$ has $\VP$-succinct hitting sets.
Then there exists a $d(n) \in \poly(n)$ such that for any functions $D(N), S(N) \in \poly(N)$, there exists a size-function $s(n) \in \poly(n)$ such that $\VP_{D,S}$ has $\VP_{d,s}$-succinct hitting sets.

Since $t(n) \in n^{\omega(1)}$, there is a finite $n_0 \in \N$, beyond which any polynomial in $\VP_{d,s}(n)$ can be simulated using the ``$(d(n),t(n))$-universal circuit'' $U_n$.
This means that $U_n$ is a hitting set generator for $\VP_{D,S}(N)$ infinitely often, which implies that the family $\set{U_n}$ is a succinct hitting set generator family for $\VP_{D,S}$ as per \autoref{defn:succinct-hitting-set-generators}.

As the above argument goes through for any $s(n) \in \poly(n)$, the family of ``$(d(n),t(n))$-universal circuits'' is a succinct hitting set generator for $\VP$.
The complexity of this family follows directly from \autoref{lem:universal-circuit}.
\end{proof}

\subsection{Hardness-Randomness Connections}

% \subsubsection*{Hitting sets to hard polynomials}

% The following statement about obtaining non-trivial hitting sets 
% \begin{propositionwp}[Hardnes from Hitting Sets]\label{prop:hardness-from-HS}
%   Let $\pazocal{T}$ be a set of polynomials, and let $\pazocal{H}$ be a hitting set for $\pazocal{T}$.
%   If $h$ is a nonzero polynomial such that $h(\veca) = 0$ for all $\veca \in \pazocal{H}$, then $h \not\in \pazocal{T}$.
% \end{propositionwp}

% \subsubsection*{Hard polynomials to hitting set generators}

We will need the following notion of combinatorial designs (a collection of subsets of a universe with small pairwise intersection).

\begin{definition}[Combinatorial designs]
A family of sets $\{S_1,\ldots,S_N\} \subseteq [\ell]$ is said to be an $(\ell,m,n)$-design if
\begin{itemize}
\item $|S_i|=m$ for each $i \in [N]$
\item $|S_i \cap S_j| < n$ for any $i \neq j$. \qedhere
\end{itemize}
\end{definition}

Kabanets and Impagliazzo \cite{KI04} obtain hitting set generators from polynomials that are hard to compute for algebraic circuits.
The following lemma is crucial to the proof of \autoref{thm:lb-on-equations-for-vnp-slice}.

\begin{lemmawp}[HSG from Hardness \cite{KI04}]
\label{lem:KI-HSG-from-hardness}
Let $\{S_1,\ldots,S_N\}$ be an $(\ell,m,n)$-design and $f(\vecx_m)$ be an $m$-variate, individual degree $d$ polynomial that requires circuits of size $s$.
Then for fresh variables $\vecy_{\ell}$, the polynomial map $\operatorname{KI-gen}_{(N,\ell,m,n)}(f) : \F^{\ell} \rightarrow \F^n $ given by
\begin{equation}\label{eqn:KI}
  \inparen{f(\vecy_{S_1}), \ldots, f(\vecy_{S_N})}
\end{equation}
is a hitting set generator for all $N$-variate polynomials with degree and circuit-size at most $\inparen{\frac{s^{0.1}}{N(d+1)^n}} $.
\end{lemmawp}

\subsection{Some Algebraic-Geometric Concepts}

We will also use some concepts from algebraic geometry. 
We provide some intuition for these, which should be sufficient to understand the results here.
For formal definitions of these concepts, the reader can refer to any algebraic geometry text (e.g. \cite{CLO07}).

\begin{itemize}
  \item \emph{Closed set} or \emph{Variety}: A set of points $S$ in $\C^n$ is called a closed set if there exists a finite set of $n$-variate polynomials $\set{f_1,\ldots,f_r}$ such that $S$ is exactly the set of common zeroes (roots) of $f_1,\ldots,f_r$.
  Such a set is sometimes referred to as a \emph{variety}, and we shall do that for the remainder of this section\footnote{Some works reserve the term `variety' to refer to `irreducible' closed sets; this distinction will not be important here.}.
  \item \emph{(Zariski) Closure of a set}: The closure of a set $S$ is the smallest variety that contains it.
  \item \emph{Dimension of a variety}: Some varieties are clearly a single `component'; e.g. zeroes of the polynomials $\set{z,x-y^2}$ in $\R^3$. In such a case, their dimension is the dimension of this component ($1$, in the above example).\\
  When a variety can be seen as a union of several components, its dimension is that of the component with the largest dimension. For example, the dimension of the zeroes of the set $\set{xz,yz}$ in $\R^3$ is $2$.
  \item \emph{Degree of a variety}: The degree of a variety is the maximal (but finite) number of intersections that it can have with a linear affine subspace (common zeroes of a set of linear polynomials).\\
  Here, it might be helpful to think of zeroes of a single bivariate polynomial of the form $y - f(x)$ to understand the nomenclature. 
\end{itemize}

\noindent We need the following two inequalities about the degree of intersections of varieties.

\begin{lemmawp}[{Bezout's inequality~\cite[Lemma 2.2]{HS80a}}]~\label{lem:Bezout-inequality}
  Let $V, W$ be varieties. Then the degree of their intersection $\deg(V \cap W) \leq \deg(V) \cdot \deg(W)$.
\end{lemmawp}

Any finite intersection of varieties is also a variety, and hence \autoref{lem:Bezout-inequality} can be used to bound the degree of any finite intersection.
However, when we additionally have a bound on the dimension of one of the varieties, one can prove a different bound, which is sometimes tighter.

\begin{lemma}[{\cite[Proposition 2.3]{HS80a}}]~\label{lem:inequality-degree-intersection-varieties}
  For varieties $V_1,V_2,\ldots,V_k$, we have the following.
  \[
    \deg(V_1 \cap V_2 \cap \cdots \cap V_k) \leq \deg(V_1) \cdot \inparen{\max_{i>1} \deg{V_i}}^{\dim V_1}
  \]
\end{lemma}
\begin{proof}[Proof sketch]
  The proof is by induction. The base case $k=1$ is trivial, so assume it is true for $k-1$. 
  For simplicity, suppose that $V_1$ is irreducible.
  Then either $V_1 \cap V_2 = V_1$, which takes us back to the $(k-1)$ case.
  Or, $W = V_1 \cap V_2$ has dimension that is at most $(\dim V_1 - 1)$ and degree that is at most $\deg(V_1) \cdot \deg(V_2)$, and again we can apply te induction hypothesis to $W, V_3, \cdots, V_k$.
  In the general case, we apply the lemma to all the irreducible components of $V_1$. 
\end{proof}

\section{Existence of Hitting Sets}\label{sec:hitting-sets}

\subsection{Over complex numbers}\label{subsec:hitting-sets-complexes}
We state the results for complex numbers, but they extend as is for rationals and reals.

All the ideas required to prove the following theorem exist in previous works~\cite{HS80b,HS80a}, but the statements have not been worded in that generality before, so we state it here.
We also sketch the relevant proofs for completeness.

\begin{restatable}[Universal polynomials imply hitting sets]{theorem}{HittingSetsFromUniversalMaps}\label{thm:hitting-sets-from-universal-maps}
  Let $\pazocal{C} \subseteq \C[x_1,\ldots,x_n]$ be a set of polynomials of degree at most $d$, and suppose for $N = \binom{n+d}{d}$ there is a \emph{universal} polynomial $U(y_1,\ldots,y_m)(\vecx)$ of total degree $D$ that generates all polynomials from $\pazocal{C}$.
  That is, for each $f \in \pazocal{C}$, there is an $\alpha \in \C^m$ such that $U(\vecy=\alpha)(\vecx) = f(\vecx)$.

  Then there exists a set $H \subset [10m]^n$ of size at most $(D \cdot (d+1)^2)$ which is a hitting set for $\pazocal{C}$.
\end{restatable}

The key ingredient in proving the above theorem is the following theorem from the work of Heintz and Schnorr \cite{HS80a}.
It bounds dimension and degree of the variety that contains the coefficient vectors of a set of polynomials $\pazocal{C}$ in terms of the coefficient-generating-map for $\pazocal{C}$, and we note that even though the theorem is originally stated for polynomials that are computable by small algebraic circuits, the proof only uses the properties of the coefficient generating map.
It closely follows the arguments in \cite[Lemma 1]{HS80b}.

\begin{theorem}[{Rewording of \cite[Basic Theorem 3.2]{HS80a}}]~\label{thm:HS80-degree-dimension-bound}
  Let $\pazocal{C} \subseteq \C[x_1,\ldots,x_n]$ be a set of polynomials of total degree at most $d$, and let $N = \binom{n+d}{n}$ be the length of the coefficient vectors of $\pazocal{C}$.

  Suppose that there is a polynomial map $\pazocal{Q}: \C^m \rightarrow \C^N$ given by polynomials $\set{Q_{\vece}}$, such that for each $f(\vecx) \in \pazocal{C}$, there exists an $\alpha \in \C^m$, so that $f(\vecx) = \sum_{\vece} Q_{\vece}(\alpha) \cdot \vecx^{\vece}$.
  Then for $W \subseteq \C^N$ being the closure of the set of coefficient vectors of $\pazocal{C}$, we have that:
  \begin{itemize} \itemsep0pt
    \item The dimension, $\dim W \leq m$.
    \item The degree, $\deg(W) \leq (\deg{\pazocal{Q}})^{\dim W}$, where $\deg(\pazocal{Q}) := \max_{\vece} \deg(Q_{\vece})$. \qedhere
  \end{itemize}
\end{theorem}
\begin{proof}[Proof sketch]
  That $\dim W \leq m$ follows from the fact that it is the image of an $m$-variate map.
  Now, let $H_1, H_2, \ldots, H_m$ be hyperplanes in $\C^N$ such that the intersection $W_0 := W \cap H_1 \cap H_2 \cap \cdots \cap H_m$ is a finite set of size equal to $\deg(W)$.

  Note that $\pazocal{Q}^{-1}(H_1), \pazocal{Q}^{-1}(H_2), \ldots, \pazocal{Q}^{-1}(H_m)$ are \emph{hypersurfaces} in $\C^m$; that is, they are the zeroes of multivariate polynomials in $m$ variables, and their intersection is a finite set.
  Further, each of these polynomials, and therefore the hyper-surfaces seen as varieties, have degree at most $\deg(\pazocal{Q})$.
  We can now bound the size of their intersection, say $V := \pazocal{Q}^{-1}(H_1) \cap \pazocal{Q}^{-1}(H_2) \cap \cdots \cap \pazocal{Q}^{-1}(H_m)$, using \autoref{lem:Bezout-inequality}.
  Thus, $\abs{V} \leq \prod_{i \in [m]} \deg(\pazocal{Q}^{-1}(H_i)) \leq (\deg \pazocal{Q})^m$.
  Finally, as $\pazocal{Q}$ is a function (and not a relation) from $V$ to $W_0$, we get that $\deg(W) \leq \abs{V} \leq \deg(\pazocal{Q})^m$.
\end{proof}

Heintz and Schnorr~\cite{HS80a} then use these bounds to derive the existence of hitting sets of small size and bit-complexity, as follows.

\begin{theorem}[{Rewording of \cite[Theorem 4.4]{HS80a}}]~\label{thm:HS80-hitting-set-existence}
  Let $\pazocal{C} \subseteq \C[x_1,\ldots,x_n]$ be a set of degree-$d$ polynomials, and let $\pazocal{Q}: \C^m \rightarrow \C^N$ be a polynomial map such that the set $S = \set{\cvector(f) : f \in C}$ is contained inside its image: $\pazocal{Q}(\C^m)$.

  Then for $b = \deg(\pazocal{Q}) \cdot (d+1)^2$, and $t = 10 \cdot \dim(W)$, there exists a set of $t$ points $\set{\veca_1,\ldots,\veca_t} \subseteq [b]^n$ that is a hitting set for $\pazocal{C}$.
\end{theorem}
\begin{proof}[Proof sketch]
  We work with sequences of points from $[b]^n$ of length $t$, instead of subsets of size $t$; any ``hitting sequence'' clearly corresponds to a hitting set.

  The key steps in the proof are then as follows.
  \begin{itemize}
    \item \underline{The variety of coefficient vectors of $\pazocal{C}$}: This is just a direct application of \autoref{thm:HS80-degree-dimension-bound}.
    The resulting variety $W$ has $\dim(W) \leq m = t/10$ and $\deg(W) \leq \deg(\pazocal{Q})^{m} = \deg(\pazocal{Q})^{t/10}$.

    \item \underline{The variety of ``bad sequences''}: We first work with $t$-length sequences of points from the entire space $\C^n$.
    So consider the $(nt+N)$-dimensional space, where we identify the first $n \cdot t$ coordinates with sequences of $t$ many, $n$-dimensional points, and the rest with coefficients of polynomials.
    Within this space, consider the closure $\pazocal{B}$ of the set of points that contain ``bad sequences''.
    A sequence $\veca_1,\ldots,\veca_t$ is bad, if there is a \emph{nonzero} polynomial $f \in \pazocal{C}$ satisfying $f(\veca_1) = \cdots = f(\veca_t) = 0$.
    Let $\pazocal{B} \subset \C^{nt}$ be the Zariski closure of the set of all bad sequences.
    
    Now, $\dim(\pazocal{B}) \leq t(n-1) + \dim W \leq nt - t + t/10$.
    To see this, consider a projection from $\pazocal{B}$ to the first $nt$ coordinates.
    The image of this projection --- just the bad sequences --- has dimension at most $t(n-1)$, since fixing some $(n-1)$ coordinates in each of the points would give us a univariate with finitely many choices for the last coordinate.
    Then, the dimension of the ``pre-image'' $\pazocal{B}$ can be at most the sum of the dimension of the image, and the dimension of the other coordinates, which are contained in $W$.

    In other (very informal) words, each bad sequence can essentially be ``described by at most $(n-1)t + \dim W$ complex numbers''.
    
    Further, we can also obtain a simple upper-bound on the degree of $\pazocal{B}$.
    Consider $t$ different equations, each involving $N+n$ coordinates.
    Each of these equations asserts that the point given by the $n$ coordinates is a zero of the polynomial specified by the other $N$ coordinates.
    Then, $\pazocal{B}$ is the intersection of the $t$ hyper-surfaces given by these equations, and $W$.
    Since each of these equations have degree $(d+1)$, the degree of $W$, by \autoref{lem:Bezout-inequality}, is at most $\deg(W) \cdot (d+1)^t$.

    \item \underline{Variety of bad sequences, inside the grid}: With these bounds on $\pazocal{B}$, we can now bound the number of bad sequences within $[b]^n$ from above.
    
    For each of the $nt$ coordinates, consider the polynomial $(z_{i,j} - 1)(z_{i,j} - 2) \cdots (z_{i,j} - b)$ of degree $b$ with $i \in [t], j \in [n]$, and define its variety --- a hyper-surface --- $V_{i,j}$.
    Further, let $\pazocal{B}' := \pazocal{B} \cap V_{1,1} \cap \cdots \cap V_{t,n}$.
    Using \autoref{lem:inequality-degree-intersection-varieties}, the degree of $B'$ is then at most
    \begin{align*}
      (\deg \pazocal{B}) \cdot b^{\dim \pazocal{B}} 
      &\leq (\deg W) \cdot (d+1)^t \cdot b^{nt + t/10 - t}, &\\
      &\leq \deg(\pazocal{Q})^{t/10} \cdot (d+1)^t \cdot b^{nt} \cdot b^{-9t/10}, &(\deg W \leq \deg(\pazocal{Q})^{t/10})\\
      &\leq \inparen{\frac{b}{(d+1)^2}}^{t/10} \cdot (d+1)^t \cdot b^{nt} \cdot b^{-9t/10}, &\inparen{\deg(\pazocal{Q}) = \frac{b}{(d+1)^2}}\\
      &\leq b^{nt} \cdot  b^{-4t/5} \cdot (d+1)^{4t/5}, & \\
      &\ll b^{nt}. &(b/(d+1) \gg 1)
    \end{align*}

    Since $\pazocal{B}'$ is a finite set, this is a bound on its size.
    Thus, most of the sequences of $t$ points within $[b]^n$ are ``hitting sequences'', and therefore give valid hitting sets for $\pazocal{C}$.\qedhere
  \end{itemize}
\end{proof}

\noindent We now prove \autoref{thm:hitting-sets-from-universal-maps} by combining \autoref{thm:HS80-hitting-set-existence} and elementary multivariate interpolation.

\HittingSetsFromUniversalMaps*
\begin{proof}
  Due to \autoref{thm:HS80-hitting-set-existence}, all that remains to be done is to obtain a coefficient-generating-map $\pazocal{Q}$ from the given universal map $U(\vecy)(\vecx)$.
  Such a map is a direct consequence of the following fact, coming from multivariate interpolation.
  \begin{claimwp}
    Given an $n$-variate, degree-$d$ polynomial $f(\vecx)$, any coefficient $\coeff_{\vece}(f)$ can be expressed as a linear combination of its evaluations on the $(d+1)^n$ points in the set $[d+1]^n$. 
  \end{claimwp}

  The polynomial map $\pazocal{Q}$ is then just an appropriately ordered sequence of linear combinations of the evaluations of $U$ over $[d+1]^n$.
  As evaluations and linear combinations do not increase the degree or the number of variables, we can invoke \autoref{thm:HS80-hitting-set-existence} with $m = m$ and $\deg(\pazocal{Q}) = \deg(U) = D$.
  Therefore, the hitting set uses $r \leq 10 m$ points from $[t]^n$ for $t \leq \deg(\pazocal{U}) \cdot (d+1)^2 \leq D \cdot (d+1)^2$, as claimed. 
\end{proof}

We now show the existence of a ``universal polynomial'' for exponential sums.
This is a fairly easy extension of \autoref{lem:universal-circuit}, as seen below.

\begin{lemma}[Universal Polynomial for Exponential Sums]\label{lem:universal-polynomial-exp-sums}
	Let $s \geq n \geq 1$ and $d \geq 0$.
  Then for $N = \binom{n+d}{n}$ there exists a polynomial $V(y_1,\ldots,y_m)(\vecx)$ with $m \leq \poly(n,d,s)$ such that:
  	\begin{itemize}
    	\item $\deg(V) \leq \poly(s)$;
    	\item for any $f \in \mathbb{C}[x_1,\ldots,x_n]$ with $\deg_{\vecx}(f)\leq d$ that can be written as an exponential sum of size $s$, there exists a vector $\veca \in \mathbb{C}^m$ such that $f(\vecx) = V(\vecy = \veca)(\vecx)$.
  	\end{itemize} 
\end{lemma}

\begin{proof}
	Suppose $f_n(\vecx)$ is an $n$-variate, degree-$d$ polynomial that can be expressed as an exponential sum of size at most $s$.
	Then by \autoref{defn:exponential-sums}, there exists an $s$-variate, degree $s$ polynomial $g_s(\vecx,\vecz) \in \ckt_{s,s}(s)$ such that $f_n$ is obtained as a sum of the polynomials given by all the $\set{0,1}$-assignments to the $\vecz$ variables, in $g_s$.

	Using \autoref{lem:universal-circuit} for number of variables, degree and size, all bounded by $s$, we get a universal circuit $U((\vecx,\vecz),\vecy) $ for $\ckt_{s,s}(s)$ with $\abs{\vecy} = m \leq s^k $ for some constant $k$.
	Furthermore, $\deg_{\vecy}(U) \leq s^k $.
  
  The universal polynomial $V(\vecy)(\vecx)$ is then just the sum of all the $2^{s-n}$ many $\set{0,1}$-assignments to the $\vecz$-variables in $U$, and hence $\deg(V) \leq \deg(U) = \poly(s)$.
\end{proof}

We are now ready to prove the existence of hitting sets for circuits and exponential sums, whose sizes and bit-lengths, grow polynomially in the sizes of the corresponding models.

\begin{lemma}[Hitting sets for efficiently computable polynomials~\cite{HS80a}]\label{lem:non-explicit-hitting-sets-complexes}
  There are constants $c$ and $e$ such that, there are (non-explicit) hitting sets $\pazocal{H}$ for $\ckt_{d,s}(n)$ (the set of all $n$-variate polynomials with degree at most $d$ that are computable by algebraic circuits of size at most $s$) with $\pazocal{H} \subset {[(nds)^c]}^n$ and $\abs{\pazocal{H}}  = (nds)^e$. 
\end{lemma}
\begin{proof}
  First, \autoref{lem:universal-circuit} provides a universal polynomial $U(\vecx,\vecy)$ for all the polynomials in the set $\ckt_{d,s}(n)$ of degree at most $(nds)^{c_1}$, and $m := \abs{\vecy} \leq (nds)^{c_1}$.
  Then, invoking \autoref{thm:hitting-sets-from-universal-maps} for this polynomial map finishes the proof.
\end{proof}

Replacing \autoref{lem:universal-circuit} in the above argument by \autoref{lem:universal-polynomial-exp-sums} then gives us an analogous statement for exponential sums.

\begin{lemmawp}[Hitting sets for efficiently definable polynomials]\label{lem:non-explicit-hitting-sets-vnp-complexes}
  There are constants $c'$ and $e'$ such that, there are (non-explicit) hitting sets $\pazocal{H}$ for $\expSum_{d,s}(n)$ with $\pazocal{H} \subset {[(nds)^{c'}]}^n$ and $\abs{\pazocal{H}}  = (nds)^{e'}$. 
\end{lemmawp}

\subsection{Finite fields: Hitting Sets for \textsf{VP} and \textsf{VNP}}\label{subsec:hitting-sets-finite-fields}

\begin{lemma}[{Folklore (cf. Forbes~\cite[Lemma 3.2.14]{F14})}]\label{lem:non-explicit-hitting-sets-finite-fields}
	Let $\F$  be a finite field with $|\F| \geq d^2$. Let $\pazocal{C}(n,d,s)$ be the class of  polynomials in $\F[x_1,\ldots, x_n]$ of degree at most $d$ that are computable by fan-in $2$ algebraic circuits of size at most $s$. Then, there is a non-explicit hitting set for $\pazocal{C}$ of size at most $\ceil{2s \cdot \inparen{\log n + 2\log s + 4}}$.
\end{lemma}

\subsection{Finite fields: Hitting Sets for \textsf{VNP}}

\begin{lemma}\label{lem:non-explicit-finite-vnp-hitting-sets}
  Let $\F$  be a finite field with $\abs{\F} \geq d^2$. Let $\pazocal{D}(n,d,s)$ be the class of  polynomials in $\F[x_1,\ldots, x_n]$ of degree at most $d$ that are $s$-definable.
  Then, there is a non-explicit hitting set $\pazocal{H}$ for $\pazocal{D}(n,d,s)$ of size at most $\ceil{2s \cdot \inparen{3\log s + 4}}$.
\end{lemma}

\begin{proof}
  In order to prove the existence of a hitting set for the class $\pazocal{D}(n,d,s)$, we will need a bound on the number of polynomials in the class $\pazocal{D}(n,d,s)$ as well as a bound on the size of an explicit hitting set for the class of  $n$-variate degree at most $d$ polynomials.
  These two bounds are summarized in the following claims, proofs of which can be found in \cite{F14}.
  %The two main statements that will be needed are the following.
  \begin{claim}[Lemma 3.1.6 in \cite{F14}]\label{clm:count-sDefinable}
    Let $\F$ be a finite field and $n, s \geq 1$. There are at most $(8n \abs{\F}s^2)^s$ $n$-variate polynomials in $\F[\vecx]$ computable by (single-output) algebraic circuits of size $\leq s$ and fan-in $\leq 2$.
  \end{claim}
  \begin{claim}[Lemma 3.2.13 in \cite{F14}]\label{clm:HSsize}
    Let $\F$ be a finite field with $\abs{\F} \geq (1 + \epsilon)d$. Let $\pazocal{C} \subseteq \F[\vecx]$ be a finite set of $n$-variate polynomials of degree $< d$. Then there is a non-explicit hitting set for $\pazocal{C}$ of size $\leq \ceil{\log_{1+\epsilon} \abs{\pazocal{C}}}$.
  \end{claim}
  Note that by definition, the number of $n$-variate polynomials that are $s$-definable is at most the number of polynomials in $\pazocal{C}(s,s,s)$; the class of $s$-variate polynomials of degree $\leq s$ computable by size $s$ algebraic circuits of fan-in $\leq 2$.
  Thus, by \autoref{clm:count-sDefinable}, $\abs{\pazocal{D}(n,d,s)} \leq (8 \abs{\F}s^3)^s$.
  
  The rest of the proof follows exactly along the lines of the proof of Lemma 3.2.14 in \cite{F14}.
  
  As $\abs{\F} \geq d^2$, we have $d \leq \abs{\F}$, and so $\abs{\F} \geq (1 + \epsilon)d$ for $(1 + \epsilon) = \sqrt{\abs{\F}}$. Thus, using $\epsilon = \sqrt{\abs{\F}} - 1$ in \autoref{clm:HSsize}, we get that there is a non-explicit hitting set $\pazocal{H}$ for $\pazocal{D}(n,d,s)$ of size at most 
  \[
    \ceil{\log_{\sqrt{\abs{\F}}} \abs{\pazocal{D}(n,d,s)}} \leq \ceil{\log_{\sqrt{\abs{\F}}} (8 \abs{\F}s^3)^s} = \ceil{s(2+2\log_{\abs{\F}} (8 s^3))} = \ceil{s(2+6\log_{\abs{\F}} (2s))}
  \]
  Finally, as $\abs{\F} \geq 2$, we have
  \[
  \abs{\pazocal{H}} \leq \ceil{s \cdot (2+6\log (2s))} = \ceil{2s \cdot (1+3\log(2s))} = \ceil{2s \cdot \inparen{3\log s + 4}}.
  \]
  This completes the proof.
\end{proof}

\section{Equations in the Bounded Coefficients setting}\label{sec:equations}

We will now prove our results about the existence of efficiently computable families of equations.
We begin with the results over finite fields, as their proofs are slightly simpler.

\subsection{Over finite fields}\label{subsec:equations-finite-fields}

\begin{theorem}[Hitting sets give equations]\label{thm:equations-from-hitting-sets-finite-fields}
  Let $\F$ be a finite field of size $q$, and let $n$ be large enough.
  For some $d \geq 1$, let $\K$ be an extension of $\F$, of size at least $d^2$.
  Let $\pazocal{C} \subseteq \F[x_1,\ldots,x_n]$ be a set of polynomials of degree at most $d$, and let $\pazocal{H} \subset \K^n$ be a hitting set for $\pazocal{C}$, of size $\abs{\pazocal{H}} = t$.

  Then, for $N = \binom{n+d}{d}$, there is an \emph{equation} $P_N(Z_1,\ldots,Z_N) \not\equiv 0$ for $\pazocal{C}'$, with $\deg(P_N)$ and $\size(P_N)$ at most $10 \cdot q \cdot N \cdot t \cdot (\log d)$.
\end{theorem}
\begin{proof}
  Let  $r_d = [\K:\F] = a \cdot \log d$, for some $a \leq 2$. Note that the elements of $\K$ can also be interpreted as vectors over $\F$ via an $\F$-linear map $\Phi:\K \rightarrow \F^{r_d} $.
  We can then define for any $i \in [r_d]$, $\Phi_i:\K \rightarrow \F$ to be its projection to the $i$-th coordinate. That is, $\Phi_i:\alpha \mapsto (\Phi(\alpha))_i$ for every $i \in [r_d]$.

  For $N = \binom{n+d}{n}$, let us index the set $[N]$ by the set $\vecx^{\leq d}$ of $n$-variate monomials of degree at most $d$.
  For a point $\veca \in \pazocal{H}$, we define the vector $\eval(\veca) \in \K^N$ as $\eval(\veca)_m = m(\veca)$ where $m \in \vecx^{\leq d}$ (that is, the $m$-th coordinate is the evaluation of the monomial $m$ at $\veca$).
  To get vectors over $\F$ instead, for each $i \in [r_n]$, we shall define $\eval(\veca)^{(i)} \in \F^N$ as $\eval(\veca)_m^{(i)} = \Phi_i(m(\veca))$.

    We are now ready to define the polynomial $P_N$.
  \begin{align*}
      P_N(z_m \;:\; m\in \vecx^{\leq d}) &:= \operatorname{OR}(\vecz) \cdot \prod_{\veca\in \pazocal{H}} \inparen{\prod_{i=1}^{r_d} \inparen{1 - \inparen{\sum_m z_m \cdot \eval(\veca)_m^{(i)}}^{|\F| - 1}}},\\
      \text{where }\operatorname{OR}(\vecz)& = \inparen{1 - \prod_{m \in \vecx^{\leq d_n}}\inparen{1 - z_m^{\abs{\F}-1}}}
  \end{align*}

  \paragraph{Constructibility:}
  Note that $\deg(P_N) \leq \abs{\F}\cdot \inparen{N + (\abs{\pazocal{H}} \cdot r_d)} \leq q \cdot (N + t \cdot 2 \log d)$ and the above expression immediately yields a circuit for $P_N$ of size that is at most $4 \cdot q \cdot t \cdot r_d \cdot N$ for all large enough $N$.

  \paragraph{Usefulness:}
  Now consider any polynomial $f \in \pazocal{C}$, we will show that $P_N(\overline{\coeff}(f)) = 0$. 

  For any polynomial $g \in \F[x_1,\ldots, x_n]$ with $\deg(g) \leq d$, we have
    \begin{align*}
      P(\overline{\coeff}(g)) & =  \operatorname{OR}(\overline{\coeff}(g)) \cdot \prod_{\veca \in \pazocal{H}} \inparen{\prod_{i=1}^{r_d} \inparen{1 - \inparen{\sum_m \overline{\coeff}(g)_m \cdot \eval(\veca)_m^{(i)}}^{|\F| - 1}}},\\
                              & = \operatorname{OR}(\overline{\coeff}(g)) \cdot \prod_{\veca\in \pazocal{H}} \inparen{\prod_{i=1}^{r_d} \inparen{1 - \inparen{\Phi_i(g(\veca))}^{|\F| - 1}}},\\
                                & = 
                                \begin{cases}
                                  1 & \text{if $g \neq 0$ and $g(\veca) = 0$ for all $\veca \in \pazocal{H}$,}\\
                                  0 & \text{if $g = 0$ or $g(\veca) \neq 0$ for some $\veca \in \pazocal{H}$.}
                                \end{cases}
    \end{align*}

  If $f = 0$, then $\OR(\overline{\coeff}(f)) = 0$.
  Else, since $f \in \pazocal{C}$, the set $\pazocal{H}$ is a hitting set for $f_n$.
  Therefore, there is some point $\veca \in \pazocal{H}_n$ such that $f(\veca) \neq 0$. Hence, $\set{P_N}$ vanishes on the coefficient vector of every polynomial in $\pazocal{C}$.
  Thus, $P_N$ is an equation for $\pazocal{C}$.
\end{proof}

\MainThmFiniteFields*

\begin{proof}
  Fix $c$ to be an arbitrary constant; we refer to the family $\set{P^{(c)}_N}$ as $\set{P_N}$ for ease of notation.
  The family $\set{P_N}$ is defined by constructing polynomials $P_N$ for each $n$ and $N(n)$ using \autoref{thm:equations-from-hitting-sets-finite-fields}; so let $n$ be any large enough number.
  Let $d_n= n^c$, and $s_n = n^{\log n}$ (in fact, $s_n$ can be any function that is barely super-polynomial in $n$).
  Since the size of $\F$ is a constant with respect to $n$, and we need fields of sufficiently large size for invoking \autoref{lem:non-explicit-hitting-sets-finite-fields}, we work over an extension  $\K_n$ of $\F$ of size at least $n^{2c}$ and at most $\abs{\F} \cdot n^{2c}$.
  By \autoref{lem:non-explicit-hitting-sets-finite-fields}, there are hitting sets in $\K_n^n$ for $\ckt_{d_n,s_n}(n)$ of size at most $s_n^2$; let $\pazocal{H}_n$ be such a hitting set.

  We can now apply \autoref{thm:equations-from-hitting-sets-finite-fields} for $\pazocal{C} = \ckt_{d_n,s_n}(n)$, $\K = \K_n$ and $\pazocal{H} = \pazocal{H}_n$ of size $t = s_n^2$, to obtain $P_N$ that has size and degree that is at most $10\abs{\F} \cdot N \cdot s_n^2 \cdot (\log d_n) \leq N^2$.
  The family $\set{P_N}$ is therefore in $\VP^{\F}$.

  \paragraph*{Usefulness against $\VP_{d_n}$}: Let $\set{f_n} \in \VP_{d_n,t(n)}$ for some $t(n) = \poly(n)$, and $n_0 \in \N$ be large enough, and also be such that $n_0^{\log n_0} > t(n_0)$.
  Then for all $n \geq n_0$, $\pazocal{H}_n$ contains a non-root of $f_n$, and hence $P_N$ vanishes on $\cvector(f_n)$.
    
  \paragraph{A remark on the largeness:}
  From the definition of $P_N$ in the proof of \autoref{thm:equations-from-hitting-sets-finite-fields}, any nonzero $g \in \F[x_1,\ldots, x_n]^{\leq d_n}$ such that $g(\veca) =   0$ for all $\veca \in \pazocal{H}_n$ will satisfy $P_N(\overline{\coeff}(g))\neq 0$.
  If we interpret the coefficients of $g$ as indeterminates, each equation of the form $g(\veca) = 0$ introduces one homogeneous linear constraint in these $N$ indeterminates, over the extension $\K_n$.
  Each such constraint can be interpreted as $O(\log n)$ homogeneous linear constraints, over $\F$. Since $\abs{\pazocal{H}_n} \ll N$, the set of $g$'s that are not annihilated by $P_N$ form a subspace of dimension at least $N - O(\abs{\pazocal{H}_n}\log n)$.
  Thus, there are at least $\inparen{\abs{\F}^{N - O(\abs{\pazocal{H}_n}\log n )} -1 }$ many $g$'s such that $P_N(\overline{\coeff}(g))\neq 0$. 
\end{proof}

\VNPfiniteField*
\begin{proof}
  Note that the proof of \autoref{thm:finite field} did not use any properties of $\VP$ apart from the existence of hitting sets, which was given by \autoref{lem:non-explicit-hitting-sets-finite-fields}.
  It is therefore easy to see that this theorem also follows by using \autoref{lem:non-explicit-finite-vnp-hitting-sets} instead, with basically the same asymptotic behaviors for the size, degree, and even the largeness, of the family $\set{Q^{(c)}_N}$.
\end{proof}

\subsection{Over rationals/complexes}\label{subsec:equations-char-zero}

We will prove the following general statement, which essentially says that the existence of ``efficient'' (low-variate, low-degree) universal polynomials for any class $\class{C}$, yield efficient equations for the subclass of $\class{C}$ containing polynomial families with bounded coefficients.  

\begin{restatable}[Equations from universal polynomials]{theorem}{EquationsFromUniversalMaps}\label{thm:equations-from-universal-maps-char0}
  Let $\pazocal{C} \subseteq \C[x_1,\ldots,x_n]$ be a set of polynomials of degree at most $d$, and suppose for $N = \binom{n+d}{d}$ there is a \emph{universal} polynomial $U(y_1,\ldots,y_m)(\vecx)$ of total degree $D$ that generates all polynomials from $\pazocal{C}$.
  That is, for each $f \in C$, there is an $\alpha \in \C^m$ such that $U(\vecy=\alpha)(\vecx) = f(\vecx)$.

  Then for $\pazocal{C}'$ being the set of all polynomials in $\pazocal{C}$ with coefficients in $\set{-1,0,1}$, there is an \emph{equation} $P_N(Z_1,\ldots,Z_N) \not\equiv 0$ for $\pazocal{C}'$, with $\deg(P_N), \size(P_N) = \poly(N)$.
\end{restatable}

As mentioned earlier, the proof would also generalize in a straightforward manner for polynomial families in $\pazocal{C}$ whose coefficients are bounded by $N(n)$.
We state this for coefficients in $\set{-1,0,1}$ just to avoid cumbersome notation.

Above theorem follows from \autoref{thm:hitting-sets-from-universal-maps} and the following theorem, which is the technical core of our constructions over fields of characteristic zero: \autoref{thm:complexes} and \autoref{thm:vnp-complexes}.

\begin{theorem}[Equations from hitting sets]\label{thm:equations-from-hitting-sets-char0}
  Let $n$ be large enough, and let $\pazocal{C} \subseteq \C[x_1,\ldots,x_n]$ be a set of polynomials of degree at most $d$, and let $\pazocal{H} \subseteq \set{1,2,\ldots,B}^n$ be a hitting set for $\pazocal{C}$ of size $t$.

  Then, for $N = \binom{n+d}{d}$, and for $\pazocal{C}'$ being the set of all polynomials in $\pazocal{C}$ with coefficients in $\set{-1,0,1}$, there is an \emph{equation} $P_N(Z_1,\ldots,Z_N) \not\equiv 0$ for $\pazocal{C}'$, with $\deg(P_N)$ and $\size(P_N)$ at most $B^4 \cdot t \cdot N^4$.
\end{theorem}
\begin{proof}
  The proof will proceed similar to the proof of \autoref{thm:finite field}, with a careful use of the Chinese Remainder Theorem. 

  Let $\Delta = \set{-1,0,1}$.
  For $N = \binom{n+d_n}{n}$, let us index the set $[N]$ by the set $\vecx^{\leq d}$ of $n$-variate monomials of degree at most $d_n$.
  For a point $\veca \in \Z^n$, we define the vector $\eval(\veca) \in \Q^N$ as $\eval(\veca)_m = m(\veca)$ where $m \in \vecx^{\leq d_n}$ (that is, the $m$-th coordinate is the evaluation of the monomial $m$ at $\veca$).
  Therefore, for any $n$-variate polynomial $f$ of degree at most $d$, we have $f(\veca) = \inangle{\cvector(f), \eval(\veca)}$, the inner-product.

  Note that for any $n$-variate polynomial $f$ of degree at most $d$ and coefficients in $\Delta$, and any $\veca \in \pazocal{H}$, we have $\abs{f(\veca)} \leq N \cdot B^{d}$, which unfortunately is exponential in the degree $d$.
  However, we can work with some ``proxy evaluations'' by simulating Chinese Remaindering.

  For any $\veca \in \pazocal{H}$ and a positive integer $r$, define the vector $\widetilde{\eval}_r(\veca)$ as follows:
  \[
    \widetilde{\eval}_r(\veca)_m := (m(\veca) \bmod r)\quad\text{for all $m \in \vecx^{\leq d_n}$}.
  \]
  It is to be stressed that $\widetilde{\eval}_r(\veca)$ is a vector over $\Q$, whose entries are integers between $0$ and $r-1$.

  \begin{claim}
    Suppose $f$ is a polynomial with  integer coefficients, and $\veca \in \Z^n$. If $f(\veca) \neq 0$ and $\abs{f(\veca)} \leq M$, then there is some $r \leq 2 (\log M)^2$ such that
    \[
      \inangle{\cvector(f), \widetilde{\eval}_r(\veca)} \neq 0 \bmod{r}.
    \]
  \end{claim}
  
  \begin{proof}[Proof of claim]
    Let $\ell = \log (M + 1)$, note that the LCM of the set $[\ell^2]$ is at least $2^\ell > M$.
    Since $f(\veca)$ is a nonzero integer with $\abs{f(\veca)} \leq M$, by the Chinese Remainder Theorem there is some prime $r \leq \ell^2$ such that $f(\veca) \not\equiv 0 \bmod{r}$.
    \begin{align*}
        \inangle{\cvector(f), \widetilde{\eval}_r(\veca)} 
            &\equiv \inangle{\cvector(f), {\eval}_r(\veca)} \bmod{r}\\
            &\equiv f(\veca) \bmod{r} \not\equiv 0 \bmod{r}\qedhere
    \end{align*}
  \end{proof}

  Let $M = N \cdot B^{d}$ and $\ell = \log (M +1)$.
  For any $r\in [\ell^2]$, any $\veca \in \pazocal{H}$, and any $n$-variate polynomial $f$ of degree at most $d$ and coefficients from $\Delta$, we have
  \[
    \abs{\inangle{\cvector(f), \widetilde{\eval}_r(\veca)} }\leq  N \cdot \ell^2 =: R.
  \]
  We are now ready to define the polynomial family $\set{P_N}$.
  \begin{align*}
    P_N(z_m\;:\; m\in \vecx^{\leq n}) & = \OR(\vecz) \cdot \prod_{\veca\in \pazocal{H}} \prod_{r=2}^{\ell^2} Q_r\inparen{\inangle{\vecz, \widetilde{\eval}_r(\veca)}},\\
    \text{where }Q_r(x) &= \prod_{\substack{i \in [-R,\ldots, R]\\i\bmod{r}\neq 0}} (x-i),\\
    \OR(\vecz) &= 1 - \prod_{m\in \vecx^{\leq d_n}}{(1 - z_m)}
  \end{align*}

  \paragraph{Constructibility:}
  For our setting of the underlying parameters, $M \leq N \cdot B^d$ and thus $\ell \leq n \cdot d \cdot B$; and $R \leq N \cdot (ndB)^2 \leq B^2 \cdot N \log^4 N$.
  Therefore, $P_N$ is a polynomial of degree at most $B^4 \cdot t \cdot N^2 \log^9 N$.
  Moreover, the above expression also shows that $P_N$ is computable by a circuit of size at most $B^4 \cdot t \cdot N^4$.
  All these bounds hold for all large enough $n$.

  \paragraph{Usefulness:}
  Fix a polynomial $f_n \in \pazocal{C'}$.
  We need to show that $P_N(\cvector(f_n)) = 0$.
  Note that we have $\OR(\cvector(f_n)) \neq 0$ if $f_n$ is nonzero, and $0$ if $f_n = 0$.
  Hence, it suffices to  show that $P_N(\cvector(f_n)) = 0$ for nonzero $f_n$.

  Since the set $\pazocal{H}$ is a hitting set for $\pazocal{C}$, we know that $f_n(\veca) \neq 0$ for some $\veca \in \pazocal{H}$.
  Therefore, for some $r \in [\ell^2]$, we have that $\inangle{\cvector(f), \widetilde{\eval}_r(\veca)}$ is a nonzero integer in $\set{-R,\ldots, R}$ that is not divisible by $r$.
  Hence, we have
  \begin{align*}
    Q_r\inparen{\inangle{\cvector(f), \widetilde{\eval}_r(\veca)}} & = 0,\\
    \implies P_N(\cvector(f)) & = 0.
  \end{align*}
  Thus, $P_N$ is an equation for $\pazocal{C}'$.
\end{proof}

\MainThmComplex*

\begin{proof}
  Fix $c$ to be an arbitrary constant; we refer to the family $\set{P^{(c)}_N}$ as $\set{P_N}$ for ease of notation.
  The family $\set{P_N}$ is defined by constructing polynomials $P_N$ for each $n$ and $N(n)$ using \autoref{thm:equations-from-hitting-sets-char0}; so let $n$ be any large enough number.
  Let $d_n= n^c$, and $s_n = n^{\log n}$ (in fact, $s_n$ can be any function that is barely super-polynomial in $n$).
  By \autoref{lem:non-explicit-hitting-sets-complexes}, there is a hitting set $\pazocal{H}_n \subseteq [B]^n$ for $\ckt_{d_n,s_n}(n)$ of size at most $s_n^{3e}$, for $B = s_n^{3e}$ for some constant $e$.

  We can now apply \autoref{thm:equations-from-hitting-sets-char0} for $\pazocal{C} = \ckt_{d_n,s_n}(n)$, and $\pazocal{H} = \pazocal{H}_n$ of size $t = s_n^{3e}$, to obtain $P_N$ that has size and degree that is at most $s_n^{15e} \cdot N^4 \leq N^5$ for all large $n$, for our setting of $s_n$.
  The family $\set{P_N}$ is therefore in $\VP^{\Q}$.

  \paragraph*{Usefulness against $\VP_{d_n}$}: Let $\set{f_n} \in \VP_{d_n,t(n)}$ for some $t(n) = \poly(n)$, and $n_0 \in \N$ be large enough, and also be such that $n_0^{\log n_0} > t(n_0)$.
  Then for all $n \geq n_0$, $\pazocal{H}_n$ contains a non-root of $f_n$, and hence $P_N$ vanishes on $\cvector(f_n)$.

  \paragraph{A remark on the largeness:}
  From the definition of $P_N$ in the proof of \autoref{thm:equations-from-hitting-sets-char0}, any nonzero polynomial $g \in \F[x_1,\ldots, x_n]^{\leq d_n}$ such that $g(\veca) = \inangle{\cvector(g), \eval(\veca)} =  0$ for all $\veca \in \pazocal{H}_n$, will satisfy $P_N(\cvector(g))\neq 0$.
  In order to show that there are many such $g$'s with coefficients in $\set{-1,0,1}$, we use a pigeon-hole argument, which is essentially an instance of a lemma of Siegel~\cite{S14a}.
  For completeness, we include a sketch of the argument here. 

  Consider the map $\Gamma:\Z^N \rightarrow \Z^{\abs{\pazocal{H}_n}}$ defined as
  \[
      \Gamma(z_m\;:\; m\in \vecx^{\leq d_n}) := \inparen{\inangle{\vecz, \eval(\veca)}\;:\; \veca\in \pazocal{H}_n}
  \]
  The map $\Gamma$ is linear in the sense that $\Gamma(\vecz + \vecz') = \Gamma(\vecz) + \Gamma(\vecz')$.
  Consider the restriction of $\Gamma$ on just $\set{0,1}^N$; the range of $\Gamma$ under this restriction is $\set{-M,\ldots, M}^{\abs{\pazocal{H}_n}}$, where $M = N \cdot B^d$.
  Hence, by the pigeon-hole-principle there must be some $\vecb \in \set{-M,\ldots, M}^{\abs{\pazocal{H}_n}}$ with at least $2^N / (2M+1)^{\abs{\pazocal{H}_n}}$ pre-images inside $\set{0,1}^N $.
  If $\vech_0$ is any fixed preimage, then
  \[
      \setdef{\vech - \vech_0 \in \set{-1, 0, 1}^N}{\vech \in \Gamma^{-1}(\vecb) \cap \set{0,1}^N}
  \]
  are all coefficient vectors of polynomials $g \in \Z[x_1,\ldots, x_n]^{\leq d_n}$ with coefficients in $\set{-1,0,1}$ whose coefficient vectors are not zeroes of $P_N$. 
\end{proof}

It is worth mentioning that there are $3^N$ possible polynomials in $\Z[x_1,\ldots, x_n]^{\leq d_n}$ with coefficients in $\set{-1,0,1}$. The above remark on the largeness shows that there are $2^{N - q(n)}$ many polynomials $g$ such that $P_N(\cvector(g)) \neq 0$; for some $q(n) = n^{O(\log{n})}$.

\VNPboundedCoeff*
\begin{proof}
  Again, the proof of \autoref{thm:complexes} did not use any properties of $\VP$ apart from the existence of hitting sets, which was given by \autoref{lem:non-explicit-hitting-sets-complexes}.
  This theorem then follows by using \autoref{lem:non-explicit-hitting-sets-vnp-complexes} instead, with essentially the same asymptotic behaviors for the size, degree, and even the largeness, of the family $\set{Q^{(c)}_N}$.
\end{proof}

\section{\textsf{VNP}-succinct hitting sets for \textsf{VP} when Permanent is hard}\label{sec:vnp-succinct-hitting-sets-for-vp}

We will now show that if the Permanent is exponentially hard, then so are all the equations for it.
This is shown by constructing $\VNP$-succinct hitting sets for $\VP$.
We first lay down the ideas behind our construction in some detail, and then formalize those ideas in the later parts.

\paragraph{Constructing $\VNP$-succinct hitting sets for $\VP$.}
Let us assume that for some constant $\epsilon > 0$ and for all\footnote{To be more precise, we should work with this condition for ``infinitely often'' $m\in \N$ and obtain that $\VNP$ does not have efficient equations infinitely often. We avoid this technicality for the sake of simplicity and the proof continues to hold for the more precise version with suitable additional care. } $m \in \N$, $\Perm_{m}$ requires circuits of size $2^{m^{\epsilon}} $.
Kabanets and Impagliazzo~\cite{KI04} showed that for every combinatorial design $\mathbf{D}$ (a collection of subsets of a universe, with small pairwise intersection) of appropriate parameters, the map
\[
  \g_{\Perm}(\vecz) = \left(\Perm(\vecz_S)\;:\; S \in \mathbf{D}\right),
\]
where $\vecz_S$ denotes the variables in $\vecz$ restricted to the indices in $S$, is a hitting set generator for circuits of size $2^{o(m^{\epsilon})} $.
Our main goal is to construct a \emph{efficient exponential sum} $F(\vecy,\vecz)$, such that
\begin{equation}
\label{eq:polyH}
F(\vecy,\vecz) = \sum\limits_{S\in {\cal D}} \operatorname{mon}_S(\vecy) \cdot {\sf Perm}(\vecz_S)
\end{equation}
where $\operatorname{mon}_S(\vecy)$ is a \emph{bijective} map between $\mathbf{D}$ and monomials of total degree $\leq d$ in $\vecy$ variables.

By choosing parameters carefully, this would immediately imply that any equation on $N$-variables, for $N = \binom{n+d}{d}$, that vanishes on the coefficient vectors of polynomials in $\VNP_d(n)$ (which is the $n^{th}$ slice of polynomial families in $\VNP_d$) requires size `super-polynomial in $N$'.
 
To show that the polynomial $F(\vecy,\vecz)$ in \autoref{eq:polyH} has an efficient exponential sum, we use a specific combinatorial design.
For the design ${\cal D}$ obtained via Reed-Solomon codes, every set in the design can be interpreted as a univariate polynomial $g$ of appropriate degree over a finite field.
The degree of $g$ (say $\delta$) and size of the finite field (say $p$) are related to the parameters of the design ${\cal D}$.
Now,
\begin{equation}
\label{eq:polyH2}
F(\vecy,\vecz) = \sum_{\substack{g \in \F_p[v]\\ \deg(g) \leq \delta}} \inparen{\prod_{i=0}^{\delta} y_i^{g_i}} \cdot \Perm(\vecz_{S(g)}),
\end{equation}
where $(g_0,\ldots,g_{\delta})$ is the coefficient vector of the univariate polynomial $g$.
Expressing $F(\vecy,\vecz)$ in \autoref{eq:polyH2} as a small exponential sum requires us to implement the product $\inparen{\prod\limits_{i=0}^{\delta} y_i^{g_i}}$ as a polynomial when given the binary representation of coefficients $g_0,\ldots,g_{\delta}$ via a binary vector $\vect$ of appropriate length (say $r$).
This is done via the polynomial $\Mon(\vect,\vecy)$ in \autoref{subsec:monomials} in a straightforward manner.
Furthermore, we want to algebraically implement the selection $\vecz_S$ for a set $S$ in the combinatorial design when given the vector $\vecg$ that represents the polynomial $g$ corresponding to $S$.
This is implemented via the polynomial $\RSDesign(\vect,\vecz)$ in \autoref{subsec:selections}.
Finally, we have
\begin{align*}
  F(\vecy,\vecz) &= \sum_{\vect \in \set{0,1}^r} \Mon(\vect, \vecy) \cdot \Perm(\RSDesign(\vect, \vecz))
\end{align*} 
which is clearly a small exponential sum, as $\set{\Perm_p}$ is in $\VNP$ and polynomials $\Mon(\vect,\vecy)$ and $\RSDesign(\vect, \vecz)$ are efficiently computable.
We now provide rest of the details of our proof.

\subsection*{Some notation}

We will be using the following additional pieces of notation for this section.
\begin{enumerate}
  \item For a vector $\vect = (t_1,\ldots, t_r)$, we will use the shorthand $t_{i,j}^{(a)}$ to denote the variable $t_{(i \cdot a + j + 1)}$.
  This would be convenient when we consider the coordinates of $\vect$ as blocks of length $a$.
  \item For integers $a,p$, we shall use $\Mod(a,p)$ to denote the unique integer $a_p \in [0,p-1]$ such that $a_p = a\bmod{p}$. 
\end{enumerate}

As mentioned in the overview, the strategy is to convert the hitting set generator given in \eqref{eqn:KI} into a succinct hitting set generator.
Therefore, we would like to associate the coordinates of \eqref{eqn:KI} into coefficients of a suitable polynomial.
That is, using exponential sums, we would like to build a polynomial of the form
\[
  g(y_1,\ldots, y_{\ell}, z_1,\ldots, z_t) = \sum_{m \in \vecy^{\leq d}} m \cdot f(\vecz_{S_m}),
\]
with the monomials $m \in \vecy^{\leq d}$ suitably indexing into the sets of a combinatorial design. The above expression already resembles an exponential sum, and with a little care this can be made effective.
We will first show that the different components of the above expression can be made succinct using the following constructions. 

\subsection{Building monomials from exponent vectors}\label{subsec:monomials}

For $n,r \in \N$, let $a = \floor{r/n}$, and define $\Mon_{r,n}(\vect,\vecy) $ as follows. 
\[
  \Mon_{r,n}(t_1,\ldots, t_r, y_1, \ldots, y_n) = \prod_{i=0}^{n-1} \prod_{j=0}^{a-1} \inparen{t_{i,j}^{(a)} y_{i+1}^{2^{j}} + (1 - t_{i,j}^{(a)}) }
\]
The following observation is now immediate from the definition above.
\begin{observation}
  For any $(e_1,\ldots, e_n) \in \inbracket{d}^n$, we have
  \[
    \Mon_{r,n}(\Bin(e_1),\ldots, \Bin(e_n), y_1, \ldots, y_n) = y_1^{e_1} \cdots y_n^{e_n},
  \]
  where $\Bin(e)$ is the tuple corresponding to the binary representation of $e$, and $r = n \cdot \ceil{\log_2 d}$.
  Furthermore, the polynomial $\Mon_{r,n}$ is computable by an algebraic circuit of size $\poly(n,r)$.
\end{observation}

\subsection{Indexing combinatorial designs algebraically}
\label{subsec:selections}
\newcommand{\Sel}{\operatorname{Sel}}

Next, we need to effectively compute the hard polynomial $f$ on sets of variables in a combinatorial design, indexed by the respective monomials.
We will need to simulate some computations modulo a fixed prime $p$. The following claim will be helpful for that purpose.

\begin{claim}\label{claim:interpolation}
For any $i,b,p \in \N_{\geq 0}$, there exists a unique univariate polynomial $Q_{i,b,p}(v)\in \mathbb{Q}[v]$ of degree at most $b$ such that
\[
  Q_{i,b,p}(a) = \begin{cases}
    1 & \text{if $0 \leq a < b$ and $a \equiv i~(\bmod{p})$},\\
    0 & \text{if $0 \leq a < b$ and $a \not\equiv i~(\bmod{p})$}.
  \end{cases}
\]
\end{claim}

\begin{proof}
  We can define a unique univariate polynomial $Q_{i,b,p}(v)$ satisfying the conditions of the claim via interpolation to make a unique univariate polynomial take a value of $0$ or $1$ according to the conditions of the claim.
  Since there are $b$ conditions, there always exists such a polynomial of degree at most $b$.
\end{proof}

For any $n,b,p \in \N_{\geq 0}$ with $n\geq p$, define
\[
  \Sel_{n,b,p}(u_1,\ldots, u_n, v) \triangleq \sum_{i=1}^n u_i\cdot  Q_{i,b,p}(v).
\]

\begin{observation}
  For any $n,b,p\in \N_{\geq 0}$ with $n\geq p$, for any $0 \leq a < b$, we have that
  \[
    \Sel_{n,b,p}(u_1,\ldots, u_n, a) = u_{\Mod(a,p)} = u_{a\bmod{p}}
  \]
  The degree of $\Sel_{n,b,p}$ is at most $(b+1)$ and can be computed by an algebraic circuit of size $\poly(b)$. 
\end{observation}

\begin{proof}
  From the definition of the univariate polynomial $Q_{i,b,p}(v)$ of degree $b$ in \autoref{claim:interpolation}, $Q_{i,b,p}(a)$ outputs $1$ if and only if $i= a \bmod{p}$.
Hence, $\Sel_{n,b,p}(u_1,\ldots, u_n, a)$ is $u_{a\bmod{p}}$ and is of degree at most $(b+1)$.
\end{proof}

\medskip
\noindent
And finally, we choose a specific combinatorial design to instantiate \autoref{lem:KI-HSG-from-hardness} with. 

\subsection{Reed-Solomon based combinatorial designs}\label{sec:RS-designs}

For any prime $p$ and any choice of  $a \leq p$, the following is an explicit construction of a $(p^2, p, a)$-combinatorial design of size $p^a$,  defined as follows:
\begin{quote}
  With the universe $U = \F_p \times \F_p$,
  for every univariate polynomial $g(t) \in \F_p[t]$ of degree less than $a$, we add the set $S_g = \setdef{(i,g(i))}{i\in \F_p}$ to the collection. 
\end{quote}
Since any two distinct univariate polynomials of degree less than $a$ can agree on at most $a$ points, it follows that the above is indeed a $(p^2, p, a)$-design.

The advantage of this specific construction is that it can be made succinct as follows.
For $r = a \cdot \floor{\log_2 p}$, let ${t_1,\ldots,t_r}$ be variables taking values in $\{0,1\}$.
The values assigned to $\vect$-variables can be interpreted as a univariate over $\F_p$ of degree $< a$ by considering $\vect \in \{0,1\}^r$ as a matrix with $a$ rows and $\floor{\log_2p}$ columns each~\footnote{Working with $\floor{\log_2 p}$ bits (as opposed to $\ceil{\log_2 p}$) makes the proofs much simpler, and does not affect the size of the design by much.}.
The binary vector in each row represents an element in $\mathbb{F}_p$. We illustrate this with an example.
\begin{center}
  \begin{tikzpicture}[transform shape]
    
    \node (label_t) at (-3.25,0) {$\vect = $};
    \matrix [matrix of math nodes, left delimiter=(, right delimiter=)](g) at (-1.75,0) {
      1 & 1 & 1\\
      0 & 1 & 0\\
      0 & 0 & 1\\
      1 & 0 & 0\\
      0 & 1 & 1\\
    };
    \node (convert) at (0.5,0) {$\longrightarrow$};
    \matrix [matrix of math nodes, left delimiter=(, right delimiter=)] (eg) at (2.25,0) {
      7\\
      2\\
      1\\
      4\\
      2\\
    };
    \node (label_g) at (3.75,0) {$\cong g(v)$};
    \node (info1) at (0.25,-2) {For $p = 11$, $a = 5$, $g(v) = 7 + 2v + v^2 + 4v^3 + 2v^4 \in \F_{11}[v]$,};
    \node (info2) at (0.25,-2.75) {$\vect$ is a $5 \times 3$ matrix that encodes the coefficients of $g(v)$.};

\end{tikzpicture}
\end{center}

Let $\vecz$ denote the $p^2$ variables $\set{z_{1},\ldots,z_{p^2}}$, put in into a $p \times p$ matrix.
Let $S$ be a set in the Reed-Solomon based $(p^2,p,a)$-combinatorial design.
We want to implement the selection $\vecz_S$ algebraically.
In the following, we design a vector of polynomials that outputs the vector of variables $\inparen{z_{0,g(0)\bmod{p}}^{(p)}, \ldots, z_{p-1,g(p-1)\bmod{p}}^{(p)} }$.
Note that as mentioned above the polynomial $g$ can be specified via variables $t_1,\ldots,t_r$.
That is,
\begin{align*}
  \RSDesign_{p,a}(t_1,\ldots, t_r, z_{1},\ldots, z_{p^2}) & \in (\F[\vect, \vecz])^p \quad,\quad \text{for $r = a \cdot \floor{\log_2 p}$},\\
  \RSDesign_{p,a}(t_1,\ldots, t_r, z_{1},\ldots, z_{p^2})_{i+1} &= \Sel_{p, p^3, p}\inparen{z_{i,0}^{(p)},\ldots, z_{i,p-1}^{(p)}, R_{i,a,p}(\vect)},\quad\text{for each $i \in \F_p $,}\\
  \text{where }R_{i,a,p}(\vect) & = \sum_{j=0}^{a-1} \insquare{\inparen{\sum_{k=0}^{\ell_p - 1} t_{j,k}^{(\ell_p)} \cdot 2^{k}} \cdot \Mod(i^{j}, p)},\\
  \text{with }\ell_p & = \floor{\log_2 p}.
\end{align*}

\begin{observation}
  For any prime $p$, $a\leq p$, and $\vect \in \set{0,1}^r$ for $r = a \cdot \floor{\log_2p}$, we have
  \[
    \RSDesign_{p,a}(\vect, \vecz) = \inparen{z_{i,g(i)}\;:\; i \in \F_p},
  \] where $g(v) \in \F_p[v]$ is the univariate whose coefficient vector is represented by the bit-vector $\vect$.
  Furthermore, the polynomial $\RSDesign_{p,a}$ is computable by an algebraic circuit of size $\poly(p)$.
\end{observation}
\begin{proof}
  Fix some $\vect \in \set{0,1}^r$.
From the definition of $R_{i,a,p}(\vect)$, it is clear that $R_{i,a,p}(\vect)$ returns an integer $\alpha$ such that $g(i) = \alpha\bmod p$ where $\vect$ encodes the coefficients of the polynomial $g(t)$ in binary.
Furthermore, since $\Mod(i^j,p)$ is the unique integer $c \in [0,p-1]$ with $c = i^j\bmod{p}$, it also follows that $R_{i,a,p}(\vect)$ is an integer in the range $[0,p^3]$.
Hence,
  \[
    \Sel_{p,p^3,p}\inparen{z_{i,0}^{(p)},\ldots, z_{i,p-1}^{(p)}, R_{i,a,p}(\vect)} = z_{i,g(i)}
  \]
  as claimed. 
\end{proof}

\subsection{The \texorpdfstring{$\VNP$}{\sf VNP}-succinct KI generator}

We are now ready to show the $\VNP$-succinctness of the Kabanets-Impagliazzo hitting set generator family when using a hard polynomial family from $\VNP$ and Reed-Solomon based combinatorial designs.

For a prime $p$ and for the largest number $m$ such that $m^2 \leq p $, we will use $\Perm_{[p]} \in \F[\vecy_{[p]}]$ to denote $\Perm_m$ applied to the first $m^2 $ variables of $\vecy$.

We now define the polynomial $F_{n,a,p}(\vecy_{[n]},\vecz_{[p^2]}) $ as follows.
\begin{align}
\label{eqn:VNP-poly}
  F_{n,a,p}(y_1,\ldots, y_n,z_1,\ldots, z_{p^2}) &= \sum_{\vect \in \set{0,1}^r} \Mon_{r,n}(\vect, \vecy) \cdot \Perm_{[p]}(\RSDesign_{p,a}(\vect, \vecz))\\
  \text{where }r & = a \cdot \floor{\log_2 p} \nonumber
\end{align}
It is evident from the above definition that the polynomial family $\set{F_{n,a,p}(\vecy,\vecz)}_n$ is in $\VNP$, for any $p$ that is polynomially related to $n$, when seen as a polynomial only in the $\vecy$-variables, with coefficients from $\C[\vecz]$.

From the construction, we have that
\[
  F_{n,a,p}(y_1,\ldots, y_n, z_1,\ldots z_{p^2}) = \sum_{\vece} \vecy^{\vece} \cdot \Perm_{[p]}(\vecz_{S_{\vece}}),
\]
where $\set{S_{\vece}}$ is an appropriate ordering of the Reed-Solomon based $(p^2, p,a)$-combinatorial design of size $p^a$, described in \autoref{sec:RS-designs}. 

\subsection{Putting it all together}
We are now ready to show that if the Permanent is exponentially hard, then any polynomial family $\set{P_N}$ that vanishes on the coefficient vectors of all polynomials in the class $\VNP$ requires super-polynomial size to compute it.

\begin{theorem}[Conditional Hardness of Equations for {\sf VNP}]\label{thm:lb-on-equations-for-vnp-slice}
    Let $\epsilon > 0$ be a constant. Suppose, for an $m$ large enough, we have that $\Perm_{m}$ requires circuits of size $2^{m^\epsilon}$.
  
    Then there is a constant $c$, such that for $n = m^{\epsilon/4}$, any $d \leq n$ and $N = \binom{n+d}{n}$, we have that every nonzero polynomial $P(x_1,\ldots, x_N)$ of degree $\poly(N)$ that is an equation for the set $\VNP_{d}(n)$, has $\operatorname{size}(P) \geq N^{c \cdot m^{\epsilon}}$. 
  \end{theorem}

\begin{proof}
  Let $p$ be the smallest prime larger than $m^2$; we know that $p \leq 2m^2$.
  We will again use $\Perm_{[p]} \in \C[\vecy_{[p]}]$ to denote $\Perm_m$ acting on the first $m^2 $ variables of $\vecy$.
  Therefore, if $\Perm_m$ requires size $2^{m^\epsilon}$ then so does $\Perm_{[p]}$.

  Consider the polynomial $F_{n,n,p}(\vecy_{[n]}, \vecz_{[p^2]})$ defined in \eqref{eqn:VNP-poly}, which we interpret as a polynomial in $\vecy$ with coefficients in $\C[\vecz]$.
  The individual degree in $\vecy$ is at least $d$, and at most $p$.
  Let $F_{n,n,p}^{\leq d}(\vecy_{[n]},\vecz_{[p^2]})$ denote the polynomial obtained from $F_{n,n,p}$ by discarding all terms whose \emph{total degree} in $\vecy$ exceeds $d$.
  By standard homogenization arguments, it follows that $F^{\leq d}_{n,n,p} \in \VNP_d(n)$, since $F_{n,n,p}(\vecy_{[n]}, \vecz_{[p^2]})$ is efficiently computable by exponential sums.
  Therefore,
  \[
      F^{\leq d}_{n,n,p}(\vecy, \vecz) = \sum_{\deg(\vecy^\vece) \leq d} \vecy^{\vece} \cdot \Perm_{[p]}(\vecz_{S_{\vece}}),
  \]
  where $S_{\vece}$, for various $\vece$, is an appropriate indexing into a $(p^2, p, n)$-combinatorial design of size $N$.
  Since the individual degree in $\vecy$ of $F_{n,n,p}$ was at least $d$, every coefficient of $F^{\leq d}_{n,n,p}$ is $\Perm_{[p]}(\vecz_{S})$ for some $S$ in the combinatorial design.
  In other words, the coefficient vector of $F^{\leq d}_{n,n,p}$ is precisely $\operatorname{KI-gen}_{N,p^2, p, n}(\Perm_{[p]})$.

  Now suppose that $P(x_1,\dots, x_N)$ is a nonzero equation for $\VNP_d(n)$ of degree at most $N^e$ for some $e$ that is independent of $N$.
  Then, in particular, it should be zero on the coefficient vector of $F^{\leq d}_{n,n,p}(\vecy,\veca) \in \VNP_d(n)$ for all $\veca \in \C^{p^2}$.
  By the Polynomial Identity Lemma~\cite{O22,DL78,Z79,S80}, this implies that $P$ must be zero on the coefficient vector of $F^{\leq d}_{n,n,p}(\vecy,\vecz) \in (\C[\vecz])[\vecy]$, where coefficients are formal polynomials in $\C[\vecz]$. 
  Since the coefficient vector of $F^{\leq d}_{n,n,p}(\vecy, \vecz)$ is just $\operatorname{KI-gen}_{N,p^2, p, n}(\Perm_{[p]})$, the contrapositive of \autoref{lem:KI-HSG-from-hardness} gives that either $\size(P)$ or $\deg(P)$ has to be at least,
  \[
    \frac{\size(\Perm_{[p]})^{0.1}}{N \cdot 2^n} > \frac{\size(\Perm_{m})^{0.1}}{N \cdot 2^n} > \frac{2^{0.1 m^{\epsilon}}}{N \cdot 2^n}
  \]
  Since $N = \binom{n+d}{n} \leq 2^{2n} \ll 2^{m^\epsilon}$, this is at least $N^{c m^{\epsilon}}$ for some constant $c$, for all large enough $N$.
  Thus, if $\deg(P)$ was indeed at most $N^e$, then its size must be at least $N^{c m^\epsilon}$. 
 \end{proof}

  \subsubsection*{Concluding that {\sf VNP} has no efficient equations}

  \NoNaturalProofsVNP*

  \begin{proof}
    We will show that for $d(n)=n$, there is no $D(N) = \poly(N)$ for which there are $\VP_{D}$-natural proofs for the class $\VNP_{d}$.
    So suppose that $\set{P_N}$ is a family of equations for $\VNP_{d}$, that has degree $\poly(N)$.
    This means that for \emph{all large enough} $n$, and $N = \binom{n+d}{n}$, the polynomial $P_N$ vanishes on the coefficient vectors of all polynomials in $\VNP_{d}(n)$.
  
    However, \autoref{thm:lb-on-equations-for-vnp-slice} shows that for $m$ large enough, if there is a constant $\epsilon > 0$ for which we have that $\size(\Perm_m) \geq 2^{m^{\epsilon}}$, then for $n = m^{\epsilon/4}$ and any $d\leq n$, the coefficient vectors of polynomials in $\VNP_{d}(n)$ form a hitting set for all $N$-variate, degree-$\poly(N)$ polynomials that are computable by circuits of size $\poly(N)$.
    Now suppose the Permanent family is $2^{m^{\epsilon}}$-hard for a constant $\epsilon > 0$, which means that $\Perm_m$ is $2^{m^{\epsilon}}$-hard for \emph{infinitely many} $m \in \N$.
    Then using \autoref{thm:lb-on-equations-for-vnp-slice}, we can conclude that for any family $\set{P_N} \in \VP$, we must have for \emph{infinitely many} $n$ that $P_N(\cvector(f_n)) \neq 0$ for some $f_n \in \VNP_d(n)$.
    Since the choice of $\set{P_N} \in \VP$ was arbitrary, this means that there are $\VNP_d$-succinct hitting sets for $\VP$, for $d = n$.
  \end{proof}

  \vspace{1em}

\begin{remark}\rm
  Some recent results on algebraic circuit lower bounds, starting with \cite{KSS14}, involves studying families of polynomials whose monomials come from a combinatorial design.
  A natural question is whether the membership of such polynomial families in $\VNP$ (often shown via Valiant's criterion, which in turn relies on the explicitness of the underlying designs) somehow implies the $\VNP$-succinctness of the KI generator in a blackbox manner.
  We do not know if such a blackbox transformation exists.
  Nevertheless, our proof of $\VNP$-succinctness of the KI generator proceeds along similar lines but crucially relies on the fact that the underlying combinatorial designs were constructed via Reed-Solomon codes.
  In contrast to this, the $\VNP$ membership of polynomial families based on combinatorial designs via Valiant's criterion, as in \cite{KSS14}, only seems to rely on the \emph{explicitness} of the designs, and so, at least on the surface, appears to be less dependent on the precise construction of the underlying combinatorial designs.
\end{remark}

\subsection{\textsf{VNP}-succinct hitting sets for circuits with large degree}\label{sec:app-vnp-succinct-hs-for-vp_nb}

In his famous work, {\Burgisser}~\cite{B18}\footnote{The \href{https://doi.org/10.1007/s10208-002-0059-5}{original work} is from 2004, this version corrects an error in one of the proofs.} showed that approximative circuits are particularly useful in computing factors of polynomials that have small size, irrespective of their degree.

\begin{lemmawp}[{Closure under taking factors~\cite[Theorem 1.3]{B18}}]\label{lem:burgisser-factor-closure-border}
  Let $f(\vecx)$ be an $n$-variate polynomial computable by a circuit of size $s$, and suppose for some co-prime $g(\vecx), h(\vecx)$ we have that $f = g^e h$.
  Then, $g$ can be approximately computed by an algebraic circuit of size at most $t$, where $t$ is $O(\deg(g)^7 \cdot s)$. 
\end{lemmawp}

It is easy to see that the above lemma when combined with the proof of \autoref{lem:KI-HSG-from-hardness} in the work of Kabanets and Impagliazzo~\cite{KI04} yields the following statement.

\begin{lemmawp}[{HSG from Approximative Hardness}]\label{lem:hsg-from-hardness-border}
  Let $\F$ be a field of characteristic zero, and let $\set{S_1,\ldots,S_N}$ be an $(\ell,m,n)$-design and $f(\vecx_m)$ be an $m$-variate, individual degree $d$ polynomial that requires \emph{approximative} circuits of size $s$ for some $s > \deg(f)$.
  Then, for fresh variables $\vecy_{\ell}$, the polynomial map
  \[
    \operatorname{KI-gen}_{(N,\ell,m,n)}(f) : \vecy \mapsto (f(\vecy_{S_1}), \ldots, f(\vecy_{S_N}))
  \]
  is a hitting set generator for all $N$-variate polynomials with circuit-size at most $\inparen{\frac{s^{0.1}}{N(d+1)^n}}$.
\end{lemmawp}

As a result of the above, if we assume that ${\Perm_n}$ is exponentially hard even in the approximative sense, then we can rule out all efficient equations for $\VNP$, irrespective of their degree.
The proof is exactly as that of \autoref{thm:lb-on-equations-for-vnp-slice}, except for the fact that we now use \autoref{lem:hsg-from-hardness-border} instead of \autoref{lem:KI-HSG-from-hardness}, which gives us the following approximative version of \autoref{thm:lb-on-equations-for-vnp-slice}.

\begin{theoremwp}[Approximative Hardness of Equations for {\sf VNP}]\label{thm:border-lb-on-equations-for-vnp-slice}
  Let $\epsilon > 0$ be a constant. Suppose, for an $m$ large enough, we have that $\Perm_{m}$ requires \emph{approximative} circuits of size $2^{m^\epsilon}$.

  Then there is a constant $c$, such that for $n = m^{\epsilon/4}$, any $d \leq n$ and $N = \binom{n+d}{n}$, we have that every nonzero polynomial $P(x_1,\ldots, x_N)$ that is an equation for the set $\VNP_{d}(n)$, has $\operatorname{size}(P) \geq N^{c \cdot m^{\epsilon}}$. 
\end{theoremwp}

The following is now immediate, using the proof of \autoref{thm:vnp-equations-hard}.

\NoEfficientEquationsForVNP*

\section{Open questions}\label{sec:open-questions}

Some key directions that are open for further study can be categorized as follows.

\begin{description}
  \item[Disproving the existence of natural proofs for $\VP$.] This would be equivalent to proving the existence of $\VP$-succinct hitting sets for $\VP$, analogous to \autoref{thm:vnp-equations-hard}.
  The key challenge here is that we do not know any constructions of hitting sets for circuits that follow from polynomial hardness.\\
  This is necessary because coefficient-vectors of $\VP$ forming a hitting set for $\VP$ is \emph{equivalent} to the ``evaluation vectors'' of $\VP$ forming a hitting set for $\VP$.
  The recent work of Andrews and Forbes~\cite{AF22} talks about some of the challenges in constructing hitting sets with parameters similar to this.
  The question of ``algebraic cryptography'' (see e.g. \cite{AD08}) alluded to before, is also along the same lines.

  \item[Proving the existence of natural proofs.] This would be an interesting development for any circuit class for which strong lower bounds are not known.
  Of course, such a result --- unless it proves a new lower bound --- would have to rely on some believable ``easiness assumption''.\\
  A specific question could be to show that constant-free circuits (or formulas, ABPs) have efficient equations without any restrictions on coefficients.
  
  In this context, it can be shown (as a consequence of \autoref{lem:universal-succinct-hsg-for-classes}) that a version of \autoref{thm:complexes} which works for integer coefficients with large magnitudes, say $\exp((\log N)^{\log\ast N})$, will imply $\VP$-natural proofs for all of $\VP$.
  The proof strategy for \autoref{thm:complexes} gives equations with degree that is at least linear in the magnitude of the coefficients, and is therefore unlikely to be useful for this purpose.
  It would therefore be interesting to know if there are constructions that achieve a better dependence between degree and the magnitude of the coefficients.

  \item[Designing non-natural lower bound strategies.] This is a slightly vague question, in that almost any concrete and general strategy for proving lower bounds that circumvents a possible natural-proofs-barrier would be interesting.\\
  In some sense, the recent breakthrough of Limaye, Srinivasan and Tavenas~\cite{LST24} provides one such approach: reduce the lower bound question for $\pazocal{C}$ to that for some $\pazocal{C}'$, such that $\pazocal{C}'$ admits natural proofs.
  However, it is unclear if this can be pursued as a general strategy, because this additionally requires non-trivial upper bounds (from $\pazocal{C}$ to $\pazocal{C}'$).
\end{description}

\section*{Acknowledgements}

The authors from TIFR acknowledge support of the Department of Atomic Energy, Government of India, under project no. 12-R\&D-TFR-5.01-0500.

Mrinal thanks Rahul Santhanam and Ben Lee Volk for many insightful conversations about algebraic natural proofs and succinct hitting sets.

Anamay thanks Robert Andrews, Amir Shpilka and Ben Lee Volk, and the attendees of the workshop on \emph{Proof Complexity and Meta-mathematics} at the Simons Institute for discussing their insights on algebraic natural proofs.

All the authors are grateful to Ben Lee Volk, Amir Shpilka, and the anonymous reviewers of FOCS 2020, STOC 2021, and STACS 2022, for their valuable suggestions on our preliminary works (\cite{CKRST20,KRST22}) that this paper builds on.

\bibliographystyle{customurlbst/alphaurlpp}
\bibliography{masterbib/references,masterbib/crossref}

\end{document}